\documentclass[journal]{IEEEtran}
\usepackage{amsmath,amsfonts}
\usepackage{algorithmic}
\usepackage{algorithm}
\usepackage{array}
\usepackage[caption=false,font=normalsize,labelfont=sf,textfont=sf]{subfig}
\usepackage{textcomp}
\usepackage{stfloats}
\usepackage{url}
\usepackage{verbatim}
\usepackage{graphicx}
\usepackage{cite}

\usepackage{amssymb}
\usepackage{bbding}
\usepackage{colortbl}
\usepackage[table]{xcolor}
\usepackage{stfloats}
\usepackage{multirow}
\usepackage{booktabs}
\usepackage{caption}
\usepackage{hyperref}
\hypersetup{
    colorlinks=true,
    linkcolor=red,
    filecolor=blue,      
    urlcolor=blue,
    citecolor=cyan,
}
\usepackage{soul, color, xcolor}
\soulregister{\cite}7 % 注册\cite命令
\soulregister{\citep}7 % 注册\citep命令
\soulregister{\citet}7 % 注册\citet命令
\soulregister{\ref}7 % 注册\ref命令
\soulregister{\pageref}7 % 注册\pageref命令

\begin{document}

\title{Efficient HDR Reconstruction from \\ Real-World Raw Images}

\author{
Qirui~Yang,~Yihao~Liu,~Qihua~Cheng,~Huanjing~Yue,~\IEEEmembership{Senior Member, IEEE},\\~Kun~Li,~\IEEEmembership{Senior Member, IEEE},~Jingyu~Yang\IEEEauthorrefmark{1},~\IEEEmembership{Senior Member, IEEE}

\thanks{
This work is supported by the National Natural Science Foundation of China under Grants 62231018, 62171317, and 62472308.

Q. Yang, H. Yue, and J. Yang are affiliated with the School of Electrical and Information Engineering, Tianjin University, Tianjin 300072, China (e-mail: \{yangqirui, huanjing.yue, yjy\}@tju.edu.cn).

Y. Liu is affiliated with the Shanghai Artificial Intelligence Laboratory, Shanghai 200232, China (e-mail: liuyihao14@mails.ucas.ac.cn).

K Li is with the College of Intelligence and Computing, Tianjin University, Tianjin 300350, China. (e-mail: lik@tju.edu.cn).

Q. Cheng is affiliated with Shenzhen Bit Microelectronics Technology Co., Ltd, Shenzhen 518000, China (e-mail: chengqihua@microbt.com).

\IEEEauthorrefmark{1}Corresponding author: Jingyu Yang.

\par\noindent\textit{This is the authors' accepted manuscript of the paper accepted for publication in IEEE Transactions on Emerging Topics in Computational Intelligence.}

\par\noindent\textcopyright\ 2025 IEEE. Personal use of this material is permitted. Permission from IEEE must be obtained for all other uses, in any current or future media, including reprinting/republishing this material for advertising or promotional purposes, creating new collective works, for resale or redistribution to servers or lists, or reuse of any copyrighted component of this work in other works.
}}
% The paper headers
\markboth{Journal of \LaTeX\ Class Files,~Vol.~14, No.~8, August~2021}%
{Shell \MakeLowercase{\textit{et al.}}: A Sample Article Using IEEEtran.cls for IEEE Journals}

\maketitle

%%%%%% Abstract %%%%%%
\begin{abstract}
The growing prevalence of high-resolution displays on edge devices has created a pressing need for efficient high dynamic range (HDR) imaging algorithms. However, most existing HDR methods either struggle to deliver satisfactory visual quality or incur high computational and memory costs, limiting their applicability to high-resolution inputs (typically exceeding 12 megapixels). Furthermore, current HDR dataset collection approaches are often labor-intensive and inefficient. In this work, we explore a novel and practical solution for HDR reconstruction directly from raw sensor data, aiming to enhance both performance and deployability on mobile platforms. Our key insights are threefold: (1) we propose RepUNet, a lightweight and efficient HDR network leveraging structural re-parameterization for fast and robust inference; (2) we design a new computational raw HDR data formation pipeline and construct a new raw HDR dataset, RealRaw-HDR; (3) we design a plug-and-play motion alignment loss to suppress ghosting artifacts under constrained bandwidth conditions effectively. Our model contains fewer than 830K parameters and takes less than 3 ms to process an image of 4K resolution using one RTX 3090 GPU. While being highly efficient, our model also achieves comparable performance to state-of-the-art HDR methods in terms of PSNR, SSIM, and a color difference metric.

\end{abstract}


%%%%%% Main Text %%%%%%
\section{Introduction}
Most resource-constrained cameras exhibit standard dynamic range (SDR), rendering them unable to capture the full range of brightness and color information in real-world scenes. Conversely, high dynamic range (HDR) imaging seeks to encompass a significantly broader range of luminance values, compensating for color distortions and the subtle detail loss observed in SDR images. Despite the availability of dedicated hardware for directly acquiring HDR images, such equipment is typically expensive, thus limiting its practicality for most users. As a result, there has been an increasing focus on fusion-based HDR imaging methods. 

Recent methods \cite{2019Multi, 2021ADNet, 2021HDR, 2020HDR-GAN, 2019Attention} based on convolution neural networks (CNNs) \cite{Liu2024, yu2024, wang2024, hu2024} have made impressive progress in HDR reconstruction performance, thanks to their scalability and flexibility from constructing elementary building blocks like convolutional layers. However, superior performance is usually obtained at a cost of a heavy computational burden \cite{2021ADNet, 2019Attention, liu2022ghost}. Although this can be alleviated by elaborate network structures or dedicated computing engines (e.g., GPU and NPU), the hardware cost and power consumption still limit the deployment of existing deep HDR reconstruction networks. Specifically, the growing number of high-definition screens on edge devices (e.g., smartphones, security cameras, and televisions) calls for a practical HDR reconstruction solution.

On the other hand, in the image processing pipeline, HDR reconstruction is widely used in the sRGB domain. 
% This way 
Previous methods~\cite{2019Multi, 2019Deep, 2020Deep} exploit a set of sRGB images with different exposure levels to produce an HDR image, which has made rapid development in recent years. However, they tend to overlook three critical aspects. 
% Firstly, the collection of datasets. 
\textbf{\textit{1) Dataset Collection:}} 
% The existing method of acquiring the dataset follows the method of Kalantari et al \cite{2017Deep}. 
{Existing methods~\cite{chen2022repghost, Zou_Yan_Fu_2023}} follow Kalantari et al. \cite{2017Deep} to construct datasets. 
They first make the subject static and take three sets of images with different exposures, and then make the subject move twice to take dynamic images with different exposures. 
However, this process is labor-intensive and difficult to acquire on a large scale. 
% Secondly, the ISP processing speed. 
\textbf{\textit{2) ISP Processing Speed:}} 
When obtaining raw SDR images with different exposures, the ISP pipeline \cite{2021Invertible, 0End} must be performed separately on each exposure. This incurs additional memory and computational overhead and leads to lower frame rates for HDR image output. 
% Thirdly, reconstruction quality. 
\textbf{\textit{3) Reconstruction Quality:}} 
Raw images contain more delicate details of the original sensor signal that can be lost while processing sRGB images. The limitations of current HDR reconstruction methods highlight the need for further research and development in HDR imaging.

In this paper, we propose an efficient scheme for HDR reconstruction in the raw image domain. 
By analyzing the HDR image sensor system, we design a lightweight and efficient model for raw HDR reconstruction named RepUNet. RepUNet adopts reparameterization techniques and does not contain explicit computationally expensive alignment modules, such as optical flow \cite{2017Deep}, deformation convolution \cite{2021HDR}, or attention \cite{2021ADNet, yan2020deep, wang2024UIE}, which are commonly used in existing deep learning-based HDR reconstruction methods \cite{2017Deep, 2021ADNet, 2017Deep2, Nima2019Deep}. To compensate for the absence of alignment modules, we introduce a plug-and-play alignment-free and motion-aware short-exposure-first selection loss, which encourages the network to focus on local motion patterns and alleviate misalignment between short- and long-exposure images. Consequently, our approach significantly reduces hardware costs and improves the real-time performance of HDR imaging systems. 

To further promote the HDR imaging system, we investigate the HDR sensor imaging principle. We observe that changing the \textit{Gain} of the image sensor can have a similar effect as modifying the exposure time under noise-free conditions. Leveraging this insight, we design an automatic control imaging system that captures raw images with different exposures, based on a digital camera photoelectric signal conversion model. This automatic control operable system satisfies real-world scenes' dynamic range requirements, making it a practical tool for generating high-quality HDR images. The resulting RealRaw-HDR dataset includes many SDR-HDR pairs for training and evaluation. By incorporating the unique characteristics of raw images into our approach, we can achieve superior HDR reconstruction results with increased efficiency and accuracy.

Our contributions are summarized as follows:
\begin{itemize}
\item[$\bullet$] We investigate the structure reparameterizable technique for the HDR task and propose a lightweight model, RepUNet, with Topological Convolution Block (TCB). TCB can be used to improve the HDR performance of any HDR model without introducing any extra burden for inference.
\item[$\bullet$] We introduce a plug-and-play alignment-free and motion-aware short-exposure-first selection loss to mitigate ghost artifacts. 
\item[$\bullet$] We propose a novel computational photography-based pipeline for raw HDR image formation and construct a real-world raw HDR dataset, \textit{i.e.}, RealRaw-HDR.
\end{itemize}

Our contributions represent a significant step forward in raw HDR image reconstruction research, providing an effective and efficient solution for producing high-quality HDR images. Our model contains fewer than 830K parameters and takes less than 3 ms to process an image of 4K resolution using one RTX 3090 GPU. While highly efficient, our model also outperforms the state-of-the-art HDR methods by a large margin in terms of PSNR, SSIM, and a color difference metric.
%-------------------------------------------------------------------------

\section{Related Work}
\subsection{HDR Imaging}
Recently, benefiting from the fast development of deep learning techniques, training deep neural networks for effective HDR reconstruction has become increasingly popular. Many methods apply deep neural networks \cite{li2022efficient, Nima2019Deep, song2021linear, huang2021toward} to learn the production of high-quality HDR images from a set of bracketed exposure SDR images. Kalantari et al. \cite{2017Deep} proposed a CNN-based HDR approach that employs optical flow to align SDR sRGB images before network inference. Wu et al. \cite{2017Deep2} approached HDR imaging as an image translation problem without explicit motion alignment. Yan et al. \cite{2019Attention} introduced spatial attention to achieve SDR image alignment. Liu et al. \cite{2021ADNet} presented an attention-guided deformable convolutional network for multi-frame HDR imaging. Niu et al. \cite{2020HDR-GAN} proposed a multi-frame HDR imaging method based on generative adversarial learning. Liu et al. \cite{liu2022ghost} proposed a Transformer-based \cite{parmar2018image} HDR imaging method. 
SCTNet \cite{tel2023alignment} proposed an alignment-free architecture employing semantic consistency transformers. These deep learning-based approaches \cite{2021ADNet, liu2022ghost} have consistently pushed the boundaries of state-of-the-art performance. To address the need for deployment on mobile devices, Prabhakar et al. \cite{prabhakar2020towards} introduced an efficient method for generating HDR images using a bilateral guided up-sampler and exploring zero-learning for HDR reconstruction. CEN-HDR introduced a lightweight architecture designed for real-time mobile applications. 
However, these methods reconstruct HDR based on sRGB images at the end of the ISP pipeline and only train on one dataset. They overlook the large computational and storage resources the ISP pipeline requires to process bracketed exposure raw images. It also complicates the ISP system, making it challenging for resource-limited cameras to output high-quality video/images. 

To simplify the ISP system, another class of HDR reconstruction methods is based on raw image input. Google HDR+ \cite{SamuelHasinoff2023BurstPF} produced the raw HDR image by aligning and merging a burst of raw frames with the same low exposure. Nevertheless, this approach requires a complex ISP system design and takes up a lot of DDR memory. Zou et al. \cite{Zou_Yan_Fu_2023} proposed reconstructing HDR images from a single raw image and collecting a raw/HDR paired dataset. However, this dataset is not suitable for real HDR sensors. Therefore, none of the existing methods can meet the requirements of real scenarios.

\subsection{Low-level Raw Image Processing}
Due to the merits of raw data, raw-based image processing \cite{Liang_Chen_Liu_Hsu_2020, Yue_Cheng_Mao_Cao_Yang_2022} has made significant progress in recent years. The work in \cite{yue2022recaptured} first performs the demoiréing task in the raw domain and then utilizes a pre-trained ISP module to transform the result into the sRGB domain. Yang et al. \cite{yang2023learning} proposed a single-stage network empowered by feature domain adaptation to decouple the denoising and color mapping tasks in raw low-light enhancement. Zhang et al. \cite{zhang2019zoom} constructed a real-world super-resolution dataset by designing an optical zoom system and proposed a baseline network with a bilateral contextual loss. Qian et al. \cite{qian2022rethinking} solved the joint demosaicing, denoising, and super-resolution task with the raw input. Wang et al. \cite{wang2020practical} proposed a lightweight and efficient network for raw image denoising. Sharif et al. \cite{a2021beyond} proposed a new learning-based approach to tackle the challenge of joint demosaicing and denoising on image sensors. Wei et al. \cite{wei2020physics} investigated the low-light image denoising considering the image sensor photoelectric properties. Yue et al. \cite{yue2020supervised} achieved state-of-the-art raw image denoising by constructing a dynamic video dataset with noise-clean pairs. Zou et al. \cite{zou2023rawhdr} proposed a model tailor-made for Raw images, harnessing the unique features of Raw data to facilitate the Raw-to-HDR mapping.
Learning-based raw image processing has demonstrated outstanding potential for high-performance reconstruction from raw sensor data. However, acquiring paired data in the raw domain is difficult and expensive. Our work proposes a new large-scale, high-quality raw dataset and provides a data synthesis pipeline to acquire raw SDR-HDR pairs based on the HDR sensor imaging system.
%-------------------------------------------------------------------------
%-------------------------------------------------------------------------
\section{New Dataset Formation Pipeline}\label{data_pipe}
We first analyze the sensor response of the imaging system and propose a new formation pipeline for raw HDR-paired data based on the camera response model.

\subsection{Analysis of CMOS Imaging System}\label{sec:analysis}
The essence of a CMOS image sensor is photo-electric signal conversion. For a single pixel, the number of electrons $Q$ released during the light-electric signal conversion can be ideally expressed as \cite{healey1994radiometric}:
\begin{equation}\label{a}
Q = T \int_\lambda \int_x \int_y E(x,y,\lambda)S(x,y)q(\lambda) dx dy d\lambda,
\end{equation}
where $(x,y)$ represents spatial coordinates on the sensor plane, $T$ is the integration time (exposure time), $E(x,y,\lambda)$ signifies the incident spectral irradiance, $S(x,y)$ characterizes the spatial response of the collection site, and $q(\lambda)$ is defined as the ratio (electrons/Joule) of collected electrons per incident light energy for the sensor as a function of wavelength $\lambda$.

Given that $(x,y)$ in Eq. \ref{a} pertains to a single photosensory cell, we assume that each parameter remains constant concerning position. Consequently, the coordinates $(x,y)$ can be omitted \cite{horn1979calculating}:
\begin{equation}\label{b}
Q = T\overline{S}A \int_\lambda E(\lambda)q(\lambda)d\lambda,
\end{equation}
where $\overline{S}$ denotes the expected value of $S(x,y)$ within a single photosensory cell, and $A$ denotes the effective photoreceptor area of the cell.

Subsequently, the camera amplifier circuit amplifies the electrical signal, yielding the raw camera response value through analog-to-digital conversion \cite{wagdy1987effect, zhang2021rethinking}:
\begin{equation}\label{raw_response}
D = \frac{K_{\textrm{a}}Q + V_{\textrm{offset}}}{\eta}\times K_{\textrm{d}},
\end{equation}
where $K_{\textrm{a}}$ represents the Analog Gain, $K_{\textrm{d}}$ stands for the Digital Gain, and $V_{\textrm{offset}}$ accounts for the bias voltage. $\eta$ corresponds to the quantization step associated with the bit depth.
Combining Eq. \ref{b} and Eq. \ref{raw_response}, the ideal model for the optical-to-digital conversion is modeled as:
\begin{equation}\label{raw_response2}
D = \frac{K_{\textrm{a}}T\overline{S}A \int_\lambda E(\lambda)q(\lambda)d\lambda + V_{\textrm{offset}}}{\eta} \times K_{\textrm{d}},
\end{equation}
where $D$ signifies the pixel value in the raw image, $V_{\textrm{offset}}/{\eta}$ accommodates artificially introduced bias voltage to prevent output signals below 0. The raw response value of the bias voltage (i.e., black level) can be directly read out. When the dark current is $0$, or we subtract the raw response value of the bias voltage, the pixel value can be expressed as:
\begin{equation}\label{raw_response3}
D = \frac{K_{\textrm{a}}T\overline{S}A \int_\lambda E(\lambda)q(\lambda)d\lambda}{\eta} \times K_{\textrm{d}}.
\end{equation}

We observe from Eq. \ref{raw_response3} that under noise-free conditions, adjusting the gain ($K_{\textrm{a}}, K_{\textrm{d}}$) can linearly change the camera raw response value. This linear characteristic allows us to achieve an equivalent result to modifying the exposure time ($T$) by simulating the gain, thereby obtaining a set of bracketed exposure raw images. However, there is unavoidable noise in the actual imaging process. Therefore, we follow the existing denoising methods \cite{yue2020supervised, Wei_Fu_Yang_Huang_2020} and try to avoid the effect of noise as much as possible during data acquisition.

\subsection{Formation of Short- and Long-exposure Raw Pairs}
Compared to sRGB images, HDR reconstruction from raw images has the advantages of more original information, simpler ISP processing, and less computation, making it a promising paradigm to deploy in edge devices. To this end, we construct a new raw HDR dataset with SDR-HDR data pairs, named RealRaw-HDR.

\subsubsection{Data Acquisition}\label{data_acquisition} 
Based on the analysis in Sec. \ref{sec:analysis}, we find that changing the sensor digital Gain,  $K_{\textrm{d}}$, can achieve a similar luminance to adjusting the exposure time $K_{\textrm{a}}$ on the noise-free condition. Consequently, we use a top-of-the-line FUJI-FILM GFX50S II camera with a wide-aperture lens to capture high-quality raw images. The camera has 15.5 stops of dynamic range (15.5 bit) and a $51$ megapixel medium format image sensor with a pixel size of $5.3\mu m$ (The iPhone 15 Pro Max primary camera single pixel size is only $1.22\mu m$). We also set the camera ISO to 800 or below and turned on the noise reduction feature to enhance image quality. At this point, the captured raw image has a low noise level. 

Specifically, we capture two raw images ($I_1$ and $I_2$) with the same exposure settings using a high-end camera. Meanwhile, we use a human subject to simulate motion between images and trigger the shutter twice in a rapid time interval, to simulate the relative motion between short- and long-exposure images within a dual-exposure sensor. Further, to eliminate the risk of unintended camera shake, we mount the camera on a tripod and use a remote smartphone to control the shutter release. Afterward, the raw images are black level corrected, normalized, and then processed with BM3D \cite{makinen2020collaborative} to reduce noise, which obtains nearly noise-free raw images. Note that there is a small relative motion between these two raw images, which is common in multi-frame HDR reconstruction. \textit{Our dataset will be released after the acceptance of this work.}

\begin{figure*}[t]
    \centering	   
    \centering{\includegraphics[width=0.88\textwidth]{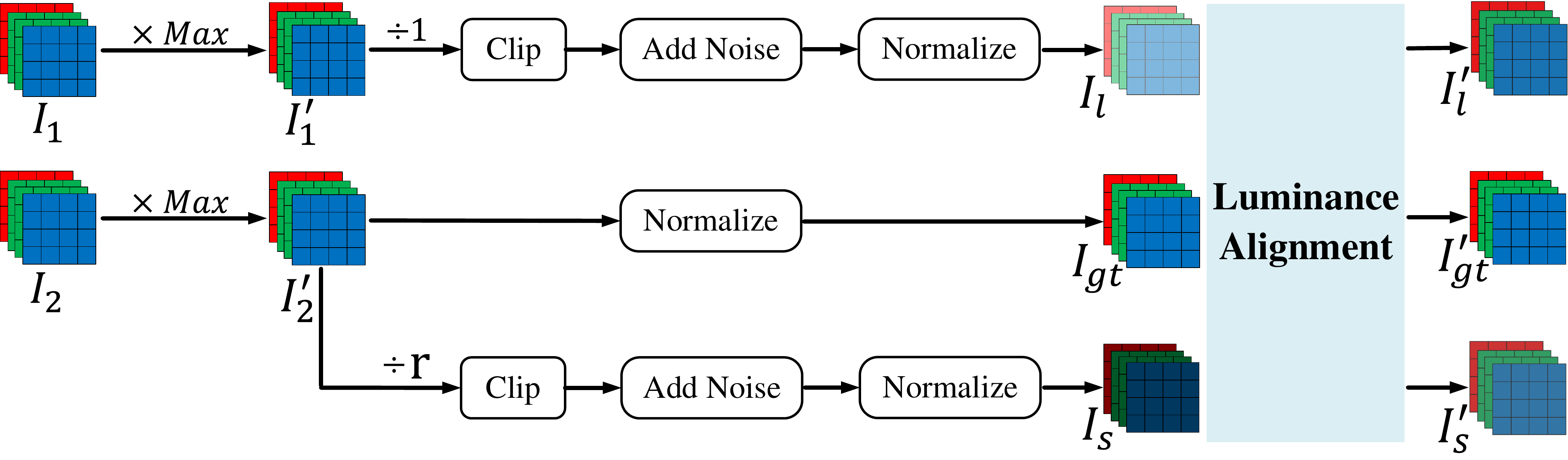}}  
    \vspace{-0.2cm}
    \caption {The raw SDR-HDR pair formation pipeline. Two clean HDR raw images, $I_1$ and $I_2$, have been processed through black-level correction and normalization. After manual digital gain, clip, add noise, and normalization, the long-exposure image $I_l$ is overexposed in the bright areas, and the short-exposure image $I_s$ has dark area information covered by noise.}
    \vspace{-0.3cm}
    \label{pipeline}
\end{figure*}

\subsubsection{Data Processing} 
Based on the digital camera imaging theory, we utilize two raw images ($I_1$ and $I_2$ have relative motion and are noise-free) to simulate short- and long-exposure images and construct the corresponding ground truths based on the principles of HDR synthesis. The Fig. \ref{pipeline} shows the proposed data formation pipeline.

\begin{figure}[t]
    \centering	   
    \centering{\includegraphics[width=8.6cm]{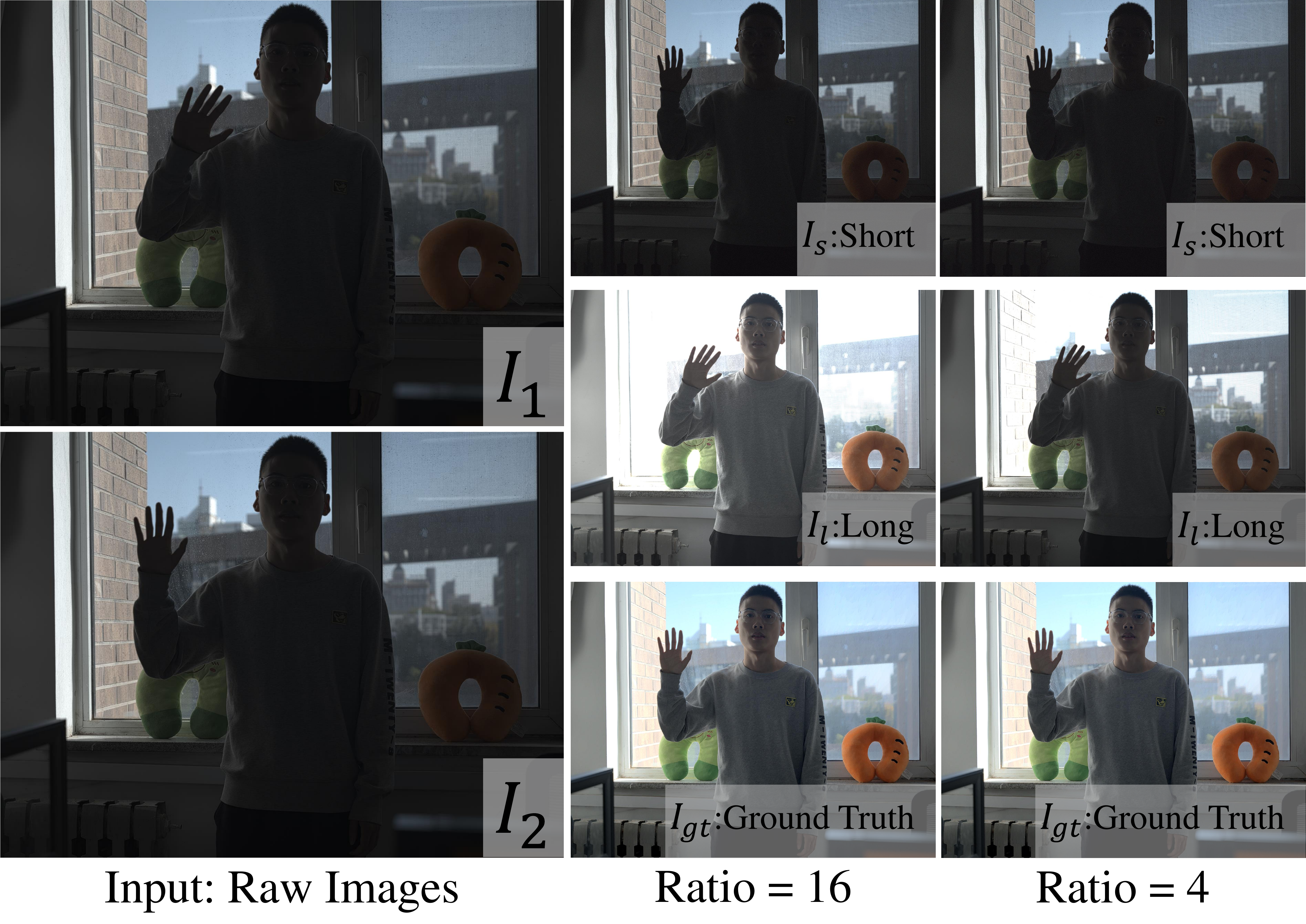}}  
    \vspace{-0.15cm}
    \caption {Real samples collected by the proposed raw SDR-HDR pair formation pipeline. For display purposes, we do not apply luminance alignment processing.}
    \vspace{-0.4cm}
    \label{example}
\end{figure}

\textbf{Selection of exposure time ratio and initial adjustment.} Commencing the pipeline, we pack two clean Bayer raw images. We select an exposure time ratio $r$ from \{4, 8, 16\} at this stage. The two normalized raw images, denoted as $I_1$ and $I_2$, are multiplied by the maximum pixel value (Max: $2^{12} \times r$). This operation yields $I^{'}_1$ and $I^{'}_2$ correspondingly. 

\textbf{Long-exposure image simulation ($I_l$).} Moving forward, we divide $I^{'}_1$ by 1 and then clip the pixel values to a range of 0 to 4095 (12 bit). This operation is equivalent to adjusting the sensor gain ($K{\textrm{d}}$), achieving an outcome comparable to altering the exposure time (as shown in Fig. \ref{example}). Then, we add noise to create the corresponding noisy long-exposure raw images. This process simulates a long-exposure noisy image ($I_l$) reaching saturation signal level (full well capacity) and the inherent noise generated by the image sensor, which preserves dark detail while losing details in brighter areas.

\textbf{Short-exposure image simulation ($I_l$).} Simultaneously, we divide $I^{'}_2$ by $r$ and clip the pixel values to a range of 0 to 4095. (Equivalent to adjusting the sensor gain). Similar to the previous step, we add noise to create the corresponding noisy short-exposure raw images. This procedure simulates a practical short-exposure image ($I_s$). This image retains detail in highlighted portions but loses darker information due to noise interference.

\textbf{Ground truth image ($I_{gt}$).} HDR aims to recover detailed information from SDR images in both brighter and darker areas. Therefore, for a dual-exposure HDR sensor, we aim to recover the darkest areas of the short-exposure image from the long-exposure image. Thus, based on this principle, we normalize $I^{'}_2$ to obtain the ground truth image $I_{gt}$. The $I_{gt}$ contains more information on bright regions than $I_l$; $I_{gt}$ has a higher signal-to-noise ratio in dark regions than $I_s$. As a result, our data formation pipeline efficiently generates an extensive array of SDR-HDR data pairs.

\textbf{Luminance alignment.} Finally, after the luminance alignment \cite{2017Deep}, we obtain the noisy raw SDR images $I^{'}_l$ and $I^{'}_s$, and the corresponding clean raw HDR image $I^{'}_{gt}$.

Our degraded dataset follows the principle of HDR synthesis—namely, the principle of maximum signal-to-noise ratio. In the darkest region, we select long-exposure images; in the brightest region, we select short-exposure images.

\subsection{RealRaw-HDR Dataset} 
\vspace{-0.1cm}
Our dataset is meticulously crafted for dual-exposure HDR sensors, supporting mainstream sensors, including Sony IMX327, IMX385, IMX585, and OV OS05B. To the best of our knowledge, there is an absence of a raw HDR dataset explicitly tailored for these HDR sensors. Our proposed data formation pipeline is efficient and user-friendly, enabling the creation of many high-quality data pairs effortlessly. We gather $240$ pairs of $8192\times6192$ high-resolution raw image pairs and expand to 720 pairs. Fig. \ref{example} shows an example of two generated SDR-HDR pairs with different exposure ratios. Additionally, by attaching an ISP pipeline to the end of our pipeline, we can create an sRGB-based HDR training dataset. In Tab. \ref{dataset}, we compare the statistics of our dataset with those of other existing HDR datasets. In this paper, all raw images have been processed with a fixed ISP, and HDR images are processed with the same tone mapping operator to obtain the sRGB version for visualization.

%-------------------------------------------------------------------------
\begin{table}[!t]
\centering
\caption{The statistics comparison between Kalanatri \cite{2017Deep}, Chen \cite{2021HDR} and our RealRaw-HDR dataset.}
\vspace{-0.1cm}
\scalebox{0.92}{
\begin{tabular}{c|c|c|c|c}
\toprule
Data & Quantity & Size & Format & Exposure Ratio \\
\midrule
Kalanatri \cite{2017Deep} & 74	& $1490\times989$	& sRGB & 4 $\&$ 8 $\&$ 16	\\ 
Chen \cite{2021HDR} & 144	& $4096\times2168$	& raw, sRGB & 4 $\&$ 8 $\&$ 16  \\
Ours & 720 & $8192\times6192$ & raw, sRGB & 4-16 \\ 
\bottomrule
\end{tabular}}
\vspace{-0.3cm}
\label{dataset}
\end{table}

\vspace{-0.1cm}
\subsection{Effectiveness of the Data Formation Pipeline}
The proposed pipeline for generating HDR data is efficient and user-friendly, allowing easy generation of numerous high-quality data pairs. Although our ground truth is derived from a single image, it contains a wide range of information characteristics of HDR images. Firstly, the raw images $I_1$ and $I_2$ are captured by an HDR camera. On the other hand, there is a significant difference in the signal-to-noise ratio between HDR images and SDR images. The exposure-aligned long-exposure images differ from the short-exposure images only in the dark and overexposed regions, in addition to the noise difference. Therefore, $I_l$ and $I_s$ have all the characteristics of real-world long and short-exposure images, and $I_{gt}$ contains a wide range of informative features of real HDR images.

\section{Methodology}   
\vspace{-0.1cm}
\subsection{Overview}\label{overview}
HDR reconstruction plays a vital role in various applications, such as mobile photography, high-definition displays, and virtual reality, where lightweight and efficient algorithms are highly demanded due to resource limitations. Previous learning-based HDR methods \cite{liu2022ghost, 2021ADNet, 2017Deep2} often rely on large and complex models, making them impractical for real-world scenarios. On the other hand, unlike GPU servers, existing optimized manipulations for mobile devices are quite limited, especially on computationally limited devices. Unsupported operators have to be processed on the CPU, which not only very low processing speed but also introduces additional MACs. Therefore, we first design a neat UNet with mobile-friendly operations as the base model, then propose a reparameterizable Topological Convolution Block to improve HDR performance. For lightweight design, we do not use the computationally demanding explicit alignment in our HDR network. To compensate for the absence of alignment modules, we introduce a plug-and-play alignment-free and motion-aware short-exposure-first selection loss (in Sec. \ref{total_loss}) that enables training with unaligned pairs. 

% \vspace{-0.3cm}
\subsection{Base Model}\label{model}
\begin{figure}[t]
\begin{minipage}[b]{0.99\linewidth}
    \centering
    \centerline{\includegraphics[width=8.6cm]{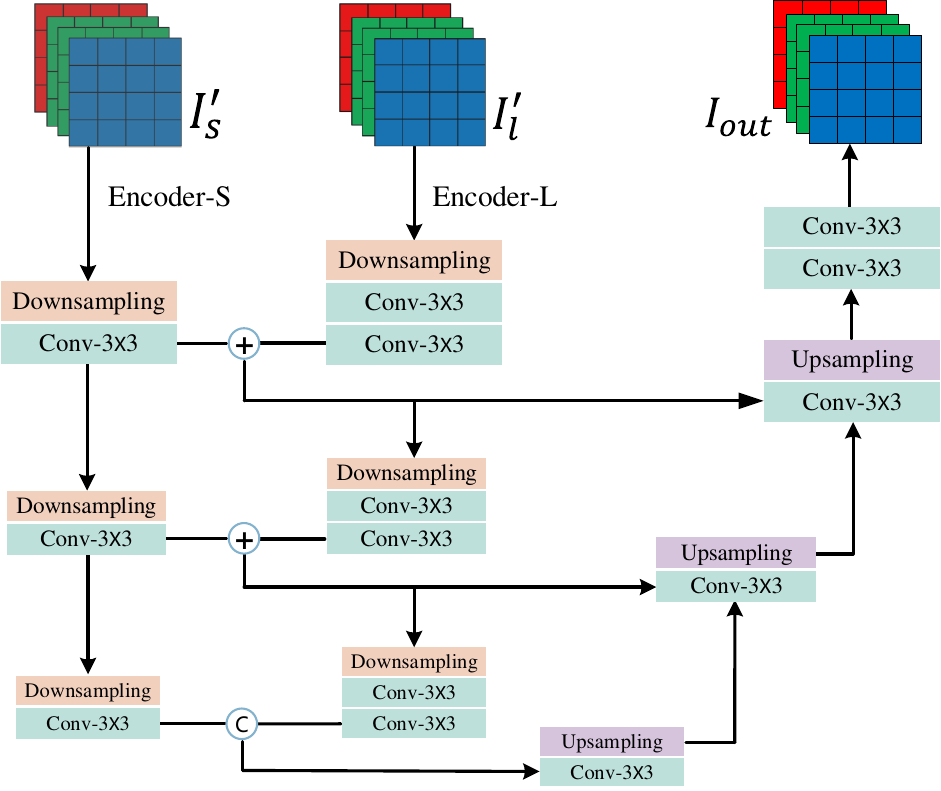}}
    \centerline{(a) Base Model}
\end{minipage} %\par
\medskip
\begin{minipage}[b]{0.99\linewidth}
    \centering
    \centerline{\includegraphics[width=8.6cm]{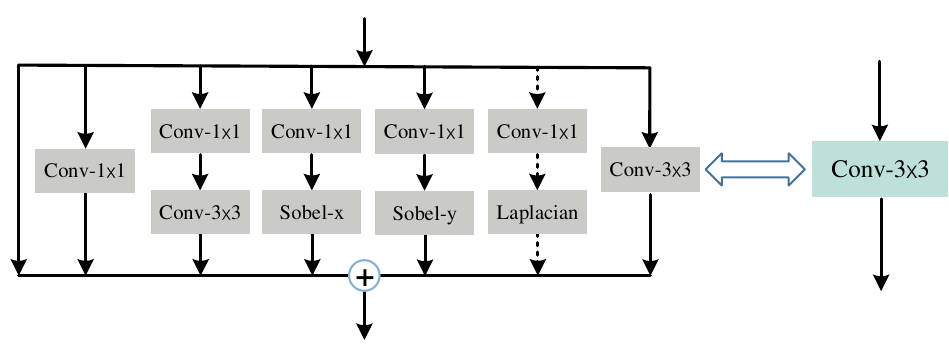}}
    \centerline{(b) Topology Convolution Block (TCB)}
\end{minipage}
\vspace{-0.4cm}
\caption{Illustration of (a) Base Model and (b) Topology Convolution Block (TCB). In the training phase, the TCB employs multiple branches, which can be merged into one normal convolution layer in the inference stage.}
\vspace{-0.3cm}
\label{arch}
\end{figure}

To pursue high inference speed on commodity mobile devices and the possibility of cross-device deployment, we carefully consider the limited computation and memory resources on mobile devices and deliberately choose a neat UNet consisting of the most basic operations as the base model. The overall architecture of the base model is shown in Fig. \ref{arch}(a), maximizing the use of the dual-exposure HDR sensor imaging characteristics. We introduce two distinct sub-encoders based on the differences in long- and short-exposure image features: Encoder-S and Encoder-L. Encoder-S extracts features from the short-exposure image, serving as reference features. In parallel, Encoder-L extracts features from the long-exposure image, offering supplementary features.

Considering the limited bandwidth, we first employ the pixel unshuffle operation \cite{gu2019self} to transfer the input raw images $I^{'}_1$ and $I^{'}_s$ from $C \times H \times W$ to $ 4C \times \frac{H}{2} \times \frac{W}{2}$ to extract multi-scale contextual information while keeping the MAC of our model as low as possible. Subsequently, reference and complementary features are extracted by different numbers of normal convolutions, respectively. To be more specific, each layer of Encoder-S consists of a pixel unshuffle $\downarrow 2$ downsampling operation, a $3 \times 3$ convolution layer, and a ReLU activation ([Down-Conv-ReLU]). At the same time, each layer of Encoder-L has the structure of [Down-Conv-ReLU-Conv-ReLU]. To promote the complementary features to learn the relative motion from the reference features, we feed the reference features of each layer to the next layer by the addition of the reference features with the complementary features. Finally, we concatenate the reference features with the complementary features and feed them to the decoder. The decoder only contains 5 normal convolutions and 3 upsampling operators. By delicate design, the proposed base model is well-suitable for mobile scenarios with high efficiency and flexibility. The network design with low MAC allows for ultra-fast inference on mobile devices, and the basic operation makes cross-device deployment easier.

\subsection{Topological Convolution Block}\label{RepUNet_network}
Although the plain base model is efficient, its HDR performance is less satisfactory compared to the complicated models, as shown in Tab. \ref{re}. We thus employ the reparameterization technique to enrich the representation capability of the base model. The reparameterization has achieved promising results on other tasks \cite{ding2021repvgg, chen2022repghost, ding2019acnet, zhang2021edge}. We design a flexible reparameterizable module called the Topological Convolution Block (TCB), which can more effectively extract edge and texture information for the HDR task. As shown in Fig. \ref{arch}(b), the TCB consists of several fundamental units:\\
(1) A standard $3\times3$ convolution for a solid foundation. The standard convolution is denoted as:
\begin{equation}
F_{n} = W_n * X + B_n,
\end{equation}
where $F_{n}$, $X$, $W_n$, and $B_n$ represent the output feature, input feature, weights, and bias of the standard convolution, respectively.
\\
(2) Extending and squeezing the convolution to enhance feature expressiveness, the expanding and squeezing feature is extracted as:
\begin{equation}
F_{es} = W_s * (W_e * X + B_e) + B_s,
\end{equation}
where {$W_e$, $B_e$} and {$W_s$, $B_s$} are the $1\times1$ expanding and $3\times3$ squeezing convolutions {weights, bias}, respectively.
\\
(3) Sobel and Laplacian operators for extracting first and second-order spatial derivatives to identify edges, i.e., using a predetermined convolution kernel to process the edge information. Denote by $D_x$ and $D_y$ the horizontal and vertical Sobel filters, and $D_{lap}$ is the Laplacian filter.
\begin{equation}
D_x = \begin{bmatrix} +1 & 0 & -1\\ +2 & 0 & -2 \\ +1 & 0 & -1 \end{bmatrix},  
\end{equation}
\begin{equation} 
D_y = \begin{bmatrix} +1 & +2 & +1\\ 0 & 0 & 0 \\ -1 & -2 & -1 \end{bmatrix}, 
\end{equation}
\begin{equation}
D_{lap} = \begin{bmatrix} 0 & +1 & 0\\ +1 & -4 & +1 \\ 0 & +1 & 0 \end{bmatrix}
\end{equation}
The combined edge information is extracted by:
\begin{equation}
F_{edge} = F_{Dx} + F_{Dy} + F_{lap},
\end{equation}
where $F_{Dx}$, $F_{Dy}$, and $F_{lap}$ represent the horizontal, vertical, 2nd-order edge information, respectively.
\\
(4) A $1\times1$ convolution to encourage information exchange between channels, denoted as:
\begin{equation}
F_{c} = W_c * X + B_c,
\end{equation}
where $F_{c}$, $W_c$, and $B_c$ represent the output feature, weights, and bias of the $1\times1$ convolution, respectively.
\\
(5) A jump connection to avoid gradient vanishing or exploding, denoted as:
\begin{equation}
F_{j} = X
\end{equation}
The output of the TCB is the combination of the five components:
\begin{equation}
F_{TCB} = F_{n} + F_{es} + F_{edge} + F_{c} + F_{j}.
\end{equation}
The combined feature map is then fed into a non-linear activation layer. PReLU is employed in our experiments. It is paramount to underscore that we exclusively employ the TCB with the Laplacian operator within the decoder. This selective approach is grounded in Laplacian operator effectiveness for noise-free images, underpinning its application to enhance feature representation in contexts devoid of noise.

\subsection{Re-parameterization for Efficient Inference}\label{Inference}
To achieve an efficient HDR network that meets the stipulated design prerequisites of low computational complexity and streamlined hardware device deployment, we simplify the TCB reparameterization into a single $3\times3$ convolution after training. Following previous works \cite{ding2021diverse, ding2019acnet, zhang2021edge}, we leverage the additivity and homogeneity of convolutions, and we merge the $1\times1$ extending and $3\times3$ squeezing convolutions into a single $3\times3$ convolution. Additionally, we combine the Sobel and Laplacian operators into a special $3\times3$ convolution with a fixed convolution kernel. The $1\times1$ convolution is achieved by padding the convolution kernel with zeros. As a result, TCB can be transformed into a $3\times3$ convolution for efficient implementation during the inference stage, as shown in Fig. \ref{arch}(b). By utilizing TCB, we achieve superior HDR results with improved efficiency.

\begin{figure}[t]
    \centering	   
    \centering{\includegraphics[width=8.6cm]{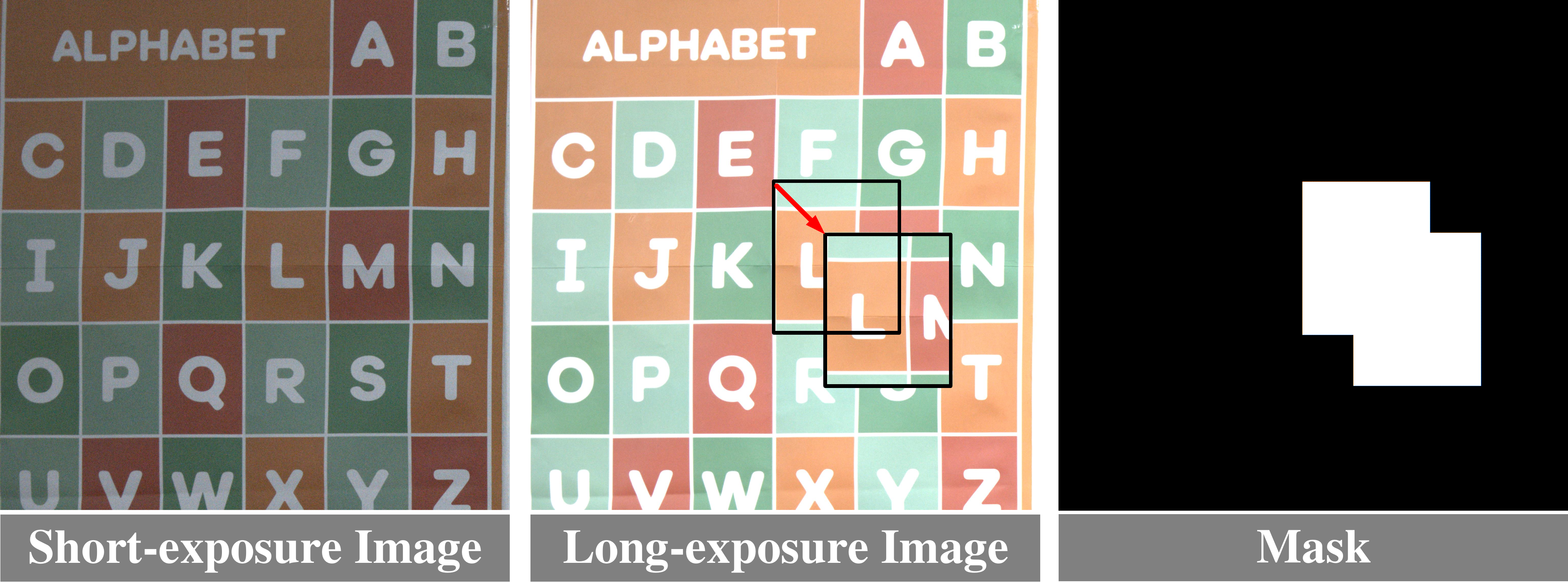}} 
    \caption {An illustrative sample of data construction for the proposed alignment-free and motion-aware short-exposure-first selection loss.}
    \vspace{-0.3cm}
    \label{ssim}
\end{figure}

\vspace{-0.2cm}
\subsection{Loss Functions}\label{total_loss} 
\textbf{Alignment-free and motion-aware short-exposure-first selection loss.} 
In fused-based HDR methods, eliminating ghosting caused by motion inconsistencies between short- and long-exposure pairs is one of the most challenging issues. Previous work \cite{2017Deep,2019Attention} commonly employs optical flow, attention mechanisms, and other methods to establish pixel correspondences between short- and long-exposure images. The objective is to suppress ghosting by designing more elaborate fusion strategies. However, these motion estimation and alignment methods are often the most computationally intensive components and cannot be accommodated by the current level of hardware design. On the other hand, unlike other image alignment tasks, such as video motion estimation and stereo matching, short- and long-exposure image fusion in HDR reconstruction does not necessarily require pixel-level correspondence. The reason is that it is challenging to recover sharp object edges due to motion blur. In contrast, short-exposure images exhibit less motion blur distortion. Therefore, ghost artifacts can be suppressed by simply detecting motion regions in long-frame images through some mechanism, discarding these pixels during the fusion process, and relying solely on the corresponding regions in short-exposure images as the exclusive information source for fusion. Based on the same consideration, overexposed regions in the long-exposure image should likewise be discarded in the fusion process. Short-exposure images are often used as reference images in engineering applications.

Based on the above analysis, we devise the strategies for dual-exposure HDR fusion: \textbf{Firstly}, in scenarios involving motion or overexposure within the fused region, we prefer to select the short-exposure image that contains more information; \textbf{Secondly}, when the SNR of the short frame is too low, we prefer to select the long-exposure image that contains more information; \textbf{Thirdly}, our strategy is inclined to address ghost artifacts with a higher priority than lower SNR when ghost artifacts and lower SNR concurrently exist.

For the designed fusion strategy, we introduce a plug-and-play alignment-free and motion-aware short-exposure-first selection loss to mitigate the ghost artifacts. We first construct a mask $M$ in the training pairs \{$I^{'}_l$, $I^{'}_s$\}. Specifically, as shown in Fig. \ref{ssim}, for each patch of an image, we randomly select a rectangle of random length and width from the long-exposure image and then move and overlap it to a random location in the range of -30 to 30 relative to the patch. The patch regions before and after the movement are labeled as 1s in the mask $M$. By introducing the mask $M$ to guide the network to focus on moving and overexposed regions, the model effectively prioritizes the short-exposure information over the long-exposed counterpart within these regions. The masks are used only in training, and the inference stage inputs only short and long exposure frames. The alignment-free and motion-aware short-exposure-first selection loss is denoted as:
\begin{equation}\label{maskloss}
L_{\textrm{AMSS}} = 1- \text{MS-SSIM}(\widetilde{I}^{out}\odot M, {I}^{gt}\odot M),
\end{equation}
where $\odot$ denotes the point-wise multiplication, and $M$ is a binary mask with ``1'' for  motion regions in long frames, and ``0'' otherwise, MS-SSIM denotes multi-scale structural similarity function. Given a mask $M$ indicating motion and overexposed regions, the above loss formula implements a strategy that encourages short-frame prioritization. 

\textbf{Reconstruction loss.} 
For saturated areas, the L2 loss punishes any deviation of pixel values from the ground truth. This allows the model to select short-exposure information in over-exposed areas. 
\begin{equation}\label{l2}
L_{\textrm{pix}} = \|\widetilde{I}^{out}-{I}^{gt}\|_2,
\end{equation}
To achieve the best HDR reconstruction results, we employ the multi-scale structural similarity loss function guide model, which learns short-exposure image information for global motion. By combining these loss functions, our model effectively produces superior results for both areas with motion and saturated regions.
\begin{equation}\label{ms}
L_{\textrm{ssim}} = 1- \text{MS-SSIM}(\widetilde{I}^{out}, {I}^{gt})
\end{equation}

\textbf{Bayer loss.}
We propose a color correction loss, named Bayer loss, to minimize color cast and artifacts. We average the two G channels of the output (RGGB pattern) and ground truth (RGGB pattern) respectively, and then concatenate the averaged G channel with the R and B channels to perform a naive transformation to the RGB color space, producing two RGB images: $\widetilde{I}^{out}_{rgb}$ and ${I}^{gt}_{rgb}$. Then, we impose the colorfulness loss between the processed output and the ground truth by the cosine embedding loss.
\begin{equation}\label{color}
L_{\textrm{b}} = \textrm{Cosine}(\widetilde{I}_{rgb}^{out}, {I}_{rgb}^{gt}),
\end{equation}
where ${\textrm{Cosine}}$ denotes cosine embedding loss \cite{he2016deep}.
The overall loss function is 
\begin{equation}\label{loss_func}
L = \alpha \cdot L_{\textrm{AMSS}} + \beta \cdot L_{\textrm{b}} + \gamma \cdot L_{\textrm{pix}}  + \eta \cdot L_{\textrm{ssim}}.
\end{equation}
where $\alpha$, $\beta$, $\gamma$, and $\eta$ are the corresponding weight coefficients.

%-------------------------------------------------------------------------
\begin{table*}[t]
\begin{center}
\caption{Performance comparison of different HDR models on synthetic dataset from RealRaw-HDR. \#Param and FLOPs represent the total number of network parameters and floating-point operations. The FLOPs and Run Times results are measured on an RTX 3090 device with a resolution of $4096 \times 2952$ raw images. Metrics with $\uparrow$ and $\downarrow$ denote higher better and lower better, respectively. PSNR test in HDR Raw images and $\Delta E$ test in tone-mapped sRGB images. The best and second-best performances are in bold and underlined, respectively. "-" indicates that inference is not possible due to memory limitations.}
\vspace{-0.1cm}
\scalebox{1}{
\begin{tabular}
{c|c|c|c|c|c|c|c|c|c|c|c}
\toprule
\multirow{2}{*}{Methods} & \multirow{2}{*}{GFLOPs} & \multirow{2}{*}{\#Param} & \multirow{2}{*}{Run Time} & \multicolumn{2}{c|}{All-Exposure} & \multicolumn{2}{c|}{Ratio=4}  & \multicolumn{2}{c|}{Ratio=8} & \multicolumn{2}{c}{Ratio=16}\\ \cline{5-12}
& & & & PSNR$\uparrow$ 	& $\Delta E$ $\downarrow$  & PSNR$\uparrow$	& $\Delta E$ $\downarrow$ & PSNR $\uparrow$	& $\Delta E\downarrow$  & PSNR$\uparrow$	& $\Delta E\downarrow$ \\
\midrule
{DeepHDR\cite{2017Deep2}}         & 2409.32 & 15.26M  & 4.3ms   & 43.3680	& 1.3767	& 43.5551  & 1.3844  & 43.7312  &1.3679  &42.8178  &1.3779\\ 
{NHDRRNet\cite{yan2020deep}}      & 826.17	 & 40.26M  & 7.9ms   & 33.0206	& 2.7308	& 33.0127  & 2.7277  & 33.0392  &2.7203  &33.0101  &2.7443\\ 
{UNet-SID\cite{chen2018learning}} & 640.89 & 7.76M  & 3.1ms    & 43.3892	& 1.3434	& 43.3314  & 1.3535  & 43.4551  &1.3312  &43.3811  &1.3456\\ 
{SGN\cite{gu2019self}}            & 712.66	 & 4.78M  & 3.3ms    & 43.6094	& 1.3235	& 43.5074  & 1.3398  & 43.7078  &1.3067  &43.6131  &1.3240	\\ 
CEN-HDR\cite{tel2022cen}         & \underline{162.35} & \textbf{0.19M}  & \textbf{2.8ms}  & 42.3021	& 1.4240	& 42.5312  & 1.4237  & 42.6531  &1.4254  &41.7219  &1.4228\\ 
HDR-GAN\cite{2020HDR-GAN}         & 629.09 & 1.32M & 8.3ms  & 44.7641	& 1.2972	& \underline{44.8954}  & \textbf{1.2867}  & 44.7743  & 1.3272  & 44.6427  & \textbf{1.2778}\\ 
{HDR-Transformer\cite{liu2022ghost}}   & 3698.28 & 1.23M  & {-}   & 44.3895	& 1.3681	& 44.3311  & 1.3811  & 44.4140  &1.3619  &44.4235  &1.3613	\\  
{AHDRNet\cite{2019Attention}}  & 2848.29 & 0.93M    & 23.6ms   & 44.7985  & \underline{1.2939} & \underline{44.8548}  & \textbf{1.2957}  & 44.8343 & \underline{1.2892} & 44.7064  & 1.2968 \\ 
SCTNet\cite{tel2023alignment}         & 3025.80 & 0.97M  & -  & \textbf{44.8452}	& 1.3048	& \textbf{44.9132}  & 1.3231  & \textbf{44.8712}  &1.3026  &\underline{44.7511}  &\underline{1.2889}\\ 
\midrule
{Ours} 	 & \textbf{127.55}  & \underline{0.82M} & \underline{2.9ms} & \underline{44.8081}  & \textbf{1.2886} & 44.7575  & \underline{1.3000}  &\underline{44.8482}  &\textbf{1.2812}  & \textbf{44.8187} &\textbf{1.2842}\\ 
\bottomrule
\end{tabular}}
\label{tab_fuji}
\vspace{-0.4cm}
\end{center}
\end{table*}

\begin{figure*}[t]
    \centering	   
    \centering{\includegraphics[width=\textwidth]{Figure/visul_fuji.pdf}}  
    \vspace{-0.5cm}
    \caption {Visual comparison of state-of-the-art HDR reconstruction methods on our synthetic dataset from RealRaw-HDR.}
    \label{visual_fuji}
\end{figure*}

\section{Experiments}
\subsection{Experimental Setup}
\textbf{Datasets and metrics.} We utilize the proposed RealRaw-HDR dataset for training. We first evaluate our method in the synthesized dataset. This test set contains 30 samples containing different exposure ratios (i.e., 4, 8, and 16) with a resolution of $4096\times2176$. To validate the validity of our method on real data, we utilize a FUJI-FILM GFX50S II camera to capture seven sets of real-world bracketed exposure raw images and the corresponding static images for generating the ground truth. Furthermore, we also utilize the Chen \cite{2021HDR} test dataset for cross-validation, which has short- and long-exposure raw pairs captured by a Sony IMX267 image sensor.

We perform a quantitative evaluation using the PSNR, SSIM, HDR-VDP3 \cite{mantiuk2023hdr}, PU \cite{azimi2021pu21}, ColorVideoVDP \cite{mantiuk2024colorvideovdp}, and CIE L*a*b* space. The HDR-VDP3 \cite{mantiuk2023hdr} metric predicts the quality degradation concerning the reference image. 
We adopt the default settings provided in the HDR-VDP-3 processing framework, which simulate a typical HDR viewing environment. Metrics are computed directly on the HDR images without any tone mapping. All HDR images are evaluated under a display profile with peak luminance of 1000 cd/m², a contrast ratio of 1000:1, a gamma of 2.2, an ambient illumination of 100 lux, and a screen size of 30 inches at $3840 \times 2160$ resolution. The viewing distance was set to 0.5 meters. 
To further validate the perceptual quality of the reconstructed HDR images, we incorporated the ColorVideoVDP \cite{mantiuk2024colorvideovdp} into our evaluation. We used the official implementation (v0.5.0) and applied it to HDR images using the configuration: "ColorVideoVDP v0.5.0, 75.4 [pix/deg], Lpeak=1000, Lblack=0.5979 [cd/m²], (standard\_4k)."
CIE L*a*b* space \footnote{CIE L*a*b* is a color space specified by the International Commission on Illumination.} describes all the colors visible to the human eye and was created to serve as a device-independent model for reference. \cite{zhang1996spatial, hill1997comparative, Liu_He_Chen_Zhang_Zhao_Dong_Qiao_2021} (also known as $\Delta E$). It offers a comprehensive quality evaluation by measuring the disparity between two HDR images within the CIE L*a*b* color space.
\begin{equation}
\Delta E = \| \widetilde{I}^{out}_{lab} - {I}^{gt}_{lab}\|_2,
\end{equation}
where $\widetilde{I}^{out}_{lab}$ and ${I}^{gt}_{lab}$ are the CIE L*a*b* version of the predicted HDR image and ground truth, respectively. Note that HDR-VDP3, PU, and $\Delta E$ are all tested on sRGB images, which are obtained from HDR Raw images through the same image signal processor process.

\begin{table*}[t]
\centering
\vspace{-0.1cm}
\caption{Performance comparison of different HDR models on the actual HDR sensor raw dataset from Chen's dataset \cite{2021HDR}. PSNR and SSIM tests in HDR Raw images. The best and second-best performances are in bold and underlined, respectively.}
\vspace{-0.15cm}
\scalebox{0.98}{
\begin{tabular}{c|c|c|c|c|c|c|c|c|c|c|c|c}
\toprule
\multirow{2}{*}{Methods} & \multicolumn{3}{c|}{All-Exposure} & \multicolumn{3}{c|}{Ratio=4} & \multicolumn{3}{c|}{Ratio=8} & \multicolumn{3}{c}{Ratio=16}\\
\cline{2-13}
            & PSNR$\uparrow$ & SSIM$\uparrow$ & $\Delta E$ $\downarrow$ & PSNR$\uparrow$   & SSIM$\uparrow$	& $\Delta E$ $\downarrow$ & PSNR$\uparrow$	 & SSIM$\uparrow$ & $\Delta E$$\downarrow$ & PSNR$\uparrow$	 & SSIM$\uparrow$ & $\Delta E$ $\downarrow$ \\
\midrule
{DeepHDR\cite{2017Deep2}}  	      & 39.4902	& 0.9731  & 2.0670	& 39.2987	& 0.9716 & 2.1201	& 40.3268	& \underline{0.9779} & 1.9159	& 38.8450	& 0.9697 & 2.1648 \\ 
{NHDRRNet\cite{yan2020deep}}      & 30.4292	& 0.9640 & 5.2132	& 30.5489	&0.9615 & 5.0771	& 30.6833	&0.9704 & 5.1679	& 30.0553	& 0.9601 & 5.3945\\ 
{UNet-SID\cite{chen2018learning}} & 39.6099	& \underline{0.9735} & 2.1527	& 39.4473	&\underline{0.9723} & 2.1860	& 40.4429	&0.9772 & 1.9640	& 38.9394	&0.9712 & 2.3081\\ 
{SGN\cite{gu2019self}}            & 39.3674	& 0.9727 & 2.3317	& 39.3531	&0.9718& 2.3357	& 40.0126	&0.9761 & 2.1956	& 38.7366	&0.9704& 2.4639\\ 
{HDR-Transformer\cite{liu2022ghost}}  & 39.9483	& 0.9726 & 2.1241	& 39.7823	&0.9715 & 2.1859	& 40.5929	&0.9750 & 2.0068	& 39.4698	&\underline{0.9713} & \underline{2.1793}\\ 
{AHDRNet\cite{2019Attention}} & \underline{40.4131} & 0.9695 & \underline{2.0123} &\textbf{40.4692}  &0.9677 & \underline{1.9829}	& \underline{41.0748}	&0.9717 & \underline{1.8519}	& \underline{39.6953}	&0.9692 & 2.2025\\ 
\midrule
{Ours} & \textbf{40.5238}	& \textbf{0.9747} & \textbf{1.9568}	& \underline{40.4061}	&\textbf{0.9733} & \textbf{1.9743}	& \textbf{41.4010}	&\textbf{0.9788} & \textbf{1.7974}	& \textbf{39.7642}	& \textbf{0.9721} & \textbf{2.0988}\\ 
\bottomrule
\end{tabular}}
% \vspace{-0.1cm}
\label{tab_chen}
\end{table*}

\begin{table*}[htb]
\centering
\caption{Performance comparison of different HDR models on the actual HDR sensor raw dataset from Chen's dataset \cite{2021HDR}. All HDR images are now evaluated under a 1000 $cd/m^2$ display profile, with metrics computed on the HDR image without tone mapping. The best and second-best performances are in bold and underlined, respectively.}
\scalebox{0.87}{
\begin{tabular}{c|c|c|c|c|c|c|c|c|c}
\toprule
\multirow{2}{*}{Methods} & \multicolumn{3}{c|}{All-Exposure} & \multicolumn{2}{c|}{Ratio=4} & \multicolumn{2}{c|}{Ratio=8} & \multicolumn{2}{c}{Ratio=16}\\
\cline{2-10}
            & HDR-VDP3$\uparrow$ & PU-PSNR $\uparrow$ & ColorVideoVDP$\uparrow $ & HDR-VDP3$\uparrow$ & PU-PSNR $\uparrow$ & HDR-VDP3$\uparrow$ & PU-PSNR $\uparrow$ & HDR-VDP3$\uparrow$ & PU-PSNR $\uparrow$ \\
\midrule
{DeepHDR\cite{2017Deep2}}  	      & 9.034  & 38.180	& 9.646 & 9.042 & 37.581	&8.939 & \underline{39.136}	& 9.121 & 37.824 \\ 
{NHDRRNet\cite{yan2020deep}}    & 8.771 & 30.743 & 9.276	&8.687 & 30.756	&8.739 & 30.946	& 8.885 & 30.567\\ 
{UNet-SID\cite{chen2018learning}} & 9.089 & 38.083 & 9.701 &9.087 & 37.613	&8.986 & 38.969	&9.194 & 37.665\\ 
{SGN\cite{gu2019self}}     & 9.104 & 37.696	& \underline{9.705} &9.074& 37.376	&9.010 & 38.368	&9.230& 37.344\\ 
{HDR-Transformer\cite{liu2022ghost}}  & 9.106 & 38.145	& 9.697 &9.093 & 37.694 &9.028 & 38.752	&9.197 & 37.988\\ 
{AHDRNet\cite{2019Attention}} & \textbf{9.153} & \underline{38.220} & \textbf{9.734} &\textbf{9.189} & \underline{37.876}	&\underline{9.037} & 38.756	&\underline{9.234} & \underline{38.038}\\ 
\midrule
{Ours} & \underline{9.138} & \textbf{39.276}	& 9.702 &\underline{9.108} & \textbf{39.459}	&\textbf{9.060} & \textbf{39.919}	& \textbf{9.247} & \textbf{38.449}\\ 
\bottomrule
\end{tabular}}
\vspace{-0.3cm}
\label{tab_vdp}
\end{table*}

\begin{figure*}[t]
    \centering	   
    \centering{\includegraphics[width=\textwidth]{Figure/visul_chen.pdf}} 
    \vspace{-0.5cm}
    \caption {Visual comparisons with the state-of-the-art methods on the actual HDR sensor raw dataset from Chen's dataset \cite{2021HDR}.}
    \vspace{-0.1cm}
    \label{visul_chen}
\end{figure*}

\textbf{Implementation details.} We train our model using the Adam optimizer \cite{DiederikPKingma2014AdamAM} with weight decay $1\times 10^{4}$, learning rate $10^{-4}$, and $\beta_{1}$ and $\beta_{2}$ values set to 0.9 and 0.999, respectively. The input patch size for the network is $256 \times 256$, and the batch size is 32. Our model is implemented in PyTorch and trained with an NVIDIA RTX 3090 GPU with Intel Xeon Platinum 8369B CPU (64 vCPUs@2.90GHz). The total training time is 60 hours.

\subsection{Comparison with the Other Methods}
We choose several representative low-level vision methods for comparisons, including six HDR methods based on sRGB images (AHDRNet \cite{2019Attention}, DeepHDR \cite{2017Deep2}, NHDRRNet \cite{yan2020deep}, HDR-GAN \cite{2020HDR-GAN}, HDR-Transformer \cite{liu2022ghost}, CEN-HDR \cite{tel2022cen}, and SCTNet \cite{tel2023alignment}), as well as two methods for denoising raw images (SGN \cite{gu2019self} and UNet-SID \cite{chen2018learning}). For fair comparisons, we retrain all the methods using the RealRaw-HDR dataset. Additionally, for AHDRNet, DeepHDR, HDR-GAN, HDR-Transformer, and NHDRRNet, we modify the network inputs to accommodate dual-exposure raw images. Similarly, for SGN and UNet-SID, we concatenate the long- and short-exposure raw pairs as inputs.

\textbf{Evaluation on the synthetic dataset.}
We first evaluated our method on a synthetic dataset generated using the raw HDR data formation pipeline. The quantitative comparison results are shown in Tab. \ref{tab_fuji}. 
The results demonstrate that our method achieves performance comparable to the state-of-the-art (SOTA) method on the synthetic dataset. Notably, our lightweight and efficient RepUNet architecture significantly reduces both model size and computational complexity, with only 0.82M parameters and 127 GFLOPs. This high efficiency enables the processing of 4K Bayer raw images in just 2.9 ms on an NVIDIA RTX 3090 GPU, outperforming other models with similar performance that require substantially longer processing times. Compared to CEN-HDR, RepUNet improves PSNR by 2.5 dB while maintaining comparable computational costs. Furthermore, RepUNet achieves performance on par with AHDRNet, SCTNet, and HDR-GAN, but with dramatically lower computational complexity: it requires only 4.5\% of the computational effort of AHDRNet (127 GFLOPs vs. 2848 GFLOPs), 4.2\% of SCTNet (127 GFLOPs vs. 2025 GFLOPs), and 20.2\% of HDR-GAN (127 GFLOPs vs. 629 GFLOPs). These results highlight the superior efficiency and practicality of our approach.
Fig. \ref{visual_fuji} shows that our method can effectively eliminate noise and ghosting artifacts in the reconstructed HDR. In comparison, DeepHDR \cite{2017Deep2}, NHDRRNet \cite{yan2020deep}, and SGN \cite{gu2019self} exhibit numerous artifacts in the palm motion region. However, our proposed RepUNet can reconstruct HDR images without ghosting (see row 2 in Fig. \ref{visual_fuji}).

\begin{figure*}[t]
    \centering	   
    \centering{\includegraphics[width=0.98\textwidth]{Figure/fujitest.pdf}}  
    \vspace{-0.1cm}
    \caption {Visual comparison of state-of-the-art HDR reconstruction methods on the FUJI Raw dataset, captured with a Fujifilm GFX50S II camera containing real-world bracketed-exposure raw images.}
    \vspace{-0.3cm}
    \label{visul_real}
\end{figure*}

\textbf{Evaluation on HDR sensor dataset.}
To validate the validity of our method on the real-world HDR sensor dataset, we utilize the Chen test dataset \cite{2021HDR} for cross-validation, which has raw images captured by a Sony IMX267 image sensor. Our method achieves comparable performance in visual quality and quantitative metrics compared to previous methods. The visual results from tests on the HDR sensor raw dataset (as shown in Fig. \ref{visul_chen}) indicate that DeepHDR, NHDRRNet, and SGN show noticeable ghosting, with NHDRRNet also suffering from color casts. Furthermore, results in Tab. \ref{tab_chen} reveal that compared to AHDRNet \cite{2019Attention}, our method yields an improvement of more than 0.35 dB and 0.05 gain in PSNR and $\Delta E$, respectively, for scenes with an exposure ratio of 8. On average, our method attains gains exceeding 0.1 dB and 0.15 in PSNR and $\Delta E$. In addition, as shown in Tab. \ref{tab_vdp}, RepUNet far outperforms the other methods in terms of PU-PSNR metrics, outperforming the suboptimal method by 1 dB. While our method yields a slightly lower score on ColorVideoVDP, it consistently outperforms competing methods on other HDR metrics (e.g., PSNR, SSIM, LPIPS), and offers lower computational cost and faster inference.

\begin{table}[t]
\centering
\vspace{-0.2cm}
\caption{Performance comparison of different HDR models on the FUJI Raw dataset, captured with a Fujifilm GFX50S II camera containing real-world bracketed-exposure raw images. The FLOPs are measured on the raw image of $7808\times5824$ resolution. The best and second-best performances are in bold and underlined, respectively.} 
\vspace{-0.1cm}
\scalebox{0.98}{
\begin{tabular}{c|c|c|c|c}
\toprule
{Methods}    & FLOPs & PSNR$\uparrow$ & SSIM$\uparrow$ & $\Delta E$$\downarrow$\\
% \cline{2-5} 
\midrule
{DeepHDR\cite{2017Deep2}}  	      & 8987.98G & 40.3610	&0.9740 & 1.5235	\\
{NHDRRNet\cite{yan2020deep}}      & 3081.94G & 33.3725 &0.9694 & 2.8789	\\
{UNet-SID\cite{chen2018learning}} & \underline{2390.83G} & 41.9495	&0.9751 & 1.4881	\\
{SGN\cite{gu2019self}}            & 2658.57G & 41.8942	&0.9750 & 1.4874	  \\
{HDR-Transformer\cite{liu2022ghost}} & 13946.42G & 41.9963	&0.9753 & 1.4897	  \\
{AHDRNet\cite{2019Attention}}     & 10625.58G & \textbf{42.6409} & \textbf{0.9763} & \textbf{1.4004} \\ 
\midrule
{Ours} & \textbf{475.83G}  & \underline{42.5364}	& \underline{0.9760} & \underline{1.4101}	\\ 
\bottomrule
\end{tabular}}
\vspace{-0.5cm}
\label{tab_real}
\end{table}

\textbf{Evaluation on FUJI raw dataset.}
We then evaluate our method on the FUJI raw datasets, which are real-world bracketed exposure raw images captured by the FUJI-FILM GFX50S II camera. Compared with previous methods, our method achieves state-of-the-art performance in visual quality and quantitative metrics. Fig. \ref{visul_real} compares results from HDR scenes, where our method achieves significantly better visualization. Our method can recover both fine details in overexposed regions and rich colors in underexposed areas without introducing artifacts (see rows 1 and 2). Compared to AHDRNet, our method can effectively remove noise and preserve the structure of dark regions. 
Tab. \ref{tab_real} shows that our method is far less computationally (FLOPs) intensive than AHDRNet \cite{2019Attention} and HDR-Transformer at similar PSNR.
Notably, the alignment module in DeepHDR, AHDRNet, HDR-Transformer, and NHDRRNet requires many line buffers, making it challenging to deploy on resource-limited edge devices. HDR-Transformer fails to perform inference even on RTX 3090 devices. In contrast, our method can alleviate ghost artifacts without relying on any alignment module and addresses color cast issues in raw images.

Although all models are trained on the RealRaw-HDR dataset, which is synthesized using the data formation pipeline, they consistently excel on both the synthetic test dataset and the real-world dataset. Particularly noteworthy is the remarkable performance achieved on the test dataset comprised of raw images captured by the HDR sensor \cite{2021HDR}. These results are solid evidence of the generalizability of our proposed RealRaw-HDR dataset and the HDR data formation pipeline. 

\vspace{-0.1cm}
\subsection{Ablation Study}
This section investigates the raw SDR-HDR pair formation pipeline and the importance of different components in the whole RepUNet. We ablate the baseline model step by step and compare the performance differences.

\textbf{Generalization of our SDR-HDR pair formation pipeline.} Our raw SDR-HDR pair formation pipeline is proposed to generate paired raw SDR-HDR data, but can also be adapted to generate paired sRGB HDR data. To demonstrate such generalization, we transform the collected RealRaw-HDR dataset with a fixed ISP pipeline into the sRGB color space, named the Raw2RGB-HDR dataset. For comparison, we train the sRGB HDR method AHDRNet \cite{2019Attention} on our Raw2RGB-HDR dataset and Kalantari dataset \cite{2017Deep} (taking the first two exposures as input, 74 pairs of images), respectively. The test dataset is from the Kalantari dataset. Results in Tab. \ref{raw-rgb} show that AHDRNet trained on our Raw2RGB-HDR dataset outperforms the one trained on the Kalantari dataset by 2.86 dB in PSNR. The performance gains benefit from an efficient and user-friendly data acquisition pipeline that generates more trainable data pairs. The results demonstrate that our data pipeline is also effective in generating paired SDR-HDR data in sRGB space. 
%%%%%%%%%%%%%%%%%%%%%%%%%%%%%
\begin{table}[htp]
\centering
\vspace{-0.2cm}
\caption{We train the HDR sRGB method AHDRNet on our Raw2RGB-HDR and Kalantari datasets, respectively. RealRGB-HDR is obtained by processing the RealRaw-HDR.}
\vspace{-0.1cm}
\scalebox{1}{
\begin{tabular}{c|c|c|c}
\toprule
Method & Dataset & PSNR & PSNR-$\mu$ \\
\midrule
\multirow{2}{*}{{AHDRNet\cite{2019Attention}}}  & Kalantari	& 35.4581	& 38.1618	\\ 
 & Raw2RGB-HDR & \textbf{38.3183} & \textbf{39.8896} \\ 
\bottomrule
\end{tabular}}
% \vspace{-0.2cm}
\label{raw-rgb}
\end{table}

\begin{table*}[t]
\centering
\vspace{-0.2cm}
\caption{Quantitative comparisons of different loss functions. AMSS-Loss represents the alignment-free and motion-aware short-exposure-first selection loss.}
\vspace{-0.2cm}
\scalebox{1}{
\begin{tabular}{c|c|c|c|c|c|c|c|c|c|c|c}
\toprule
\multirow{2}{*}{{ID}} & \multirow{2}{*}{Method} & \multirow{2}{*}{Bayer-Loss} & \multirow{2}{*}{AMSS-Loss} & \multicolumn{2}{c|}{All-Exposure} & \multicolumn{2}{c|}{Ratio=4} & \multicolumn{2}{c|}{Ratio=8} & \multicolumn{2}{c}{Ratio=16}\\
\cline{5-12}
 & & & & PSNR$\uparrow$	& $\Delta E$ $\downarrow$ & PSNR$\uparrow$	& $\Delta E$ $\downarrow$ & PSNR$\uparrow$	& $\Delta E$ $\downarrow$ & PSNR$\uparrow$	& $\Delta E\downarrow $\\
\midrule
{1} & {RepUNet} & \XSolidBrush 	 & \XSolidBrush	& 39.6403	& 2.1479	& 39.5214	& 2.1577	& 40.5216	& \underline{1.9603}	& 38.8780	&2.3256 \\ 
{2} & {RepUNet} & \Checkmark	 & \XSolidBrush	& \underline{40.0251}	& \underline{2.1162}	& \underline{39.9900}	& \underline{2.1143}	& \underline{40.8125}	& 1.9700	& \underline{39.2729}	& \underline{2.2642}\\ 
%\hline
%\rowcolor{black!10}
{3} & {RepUNet} & \Checkmark	 & \Checkmark	& \textbf{40.5238}	& \textbf{1.9568}	& \textbf{40.4061}	& \textbf{1.9743}	& \textbf{41.4010}	& \textbf{1.7974}	& \textbf{39.7642}	& \textbf{2.0988}\\ 
\bottomrule
\end{tabular}}
\label{loss}
\end{table*}

\begin{figure}[htp] 
    \centering	 
    \vspace{-0.2cm}
    \centering{\includegraphics[width=8.5cm]{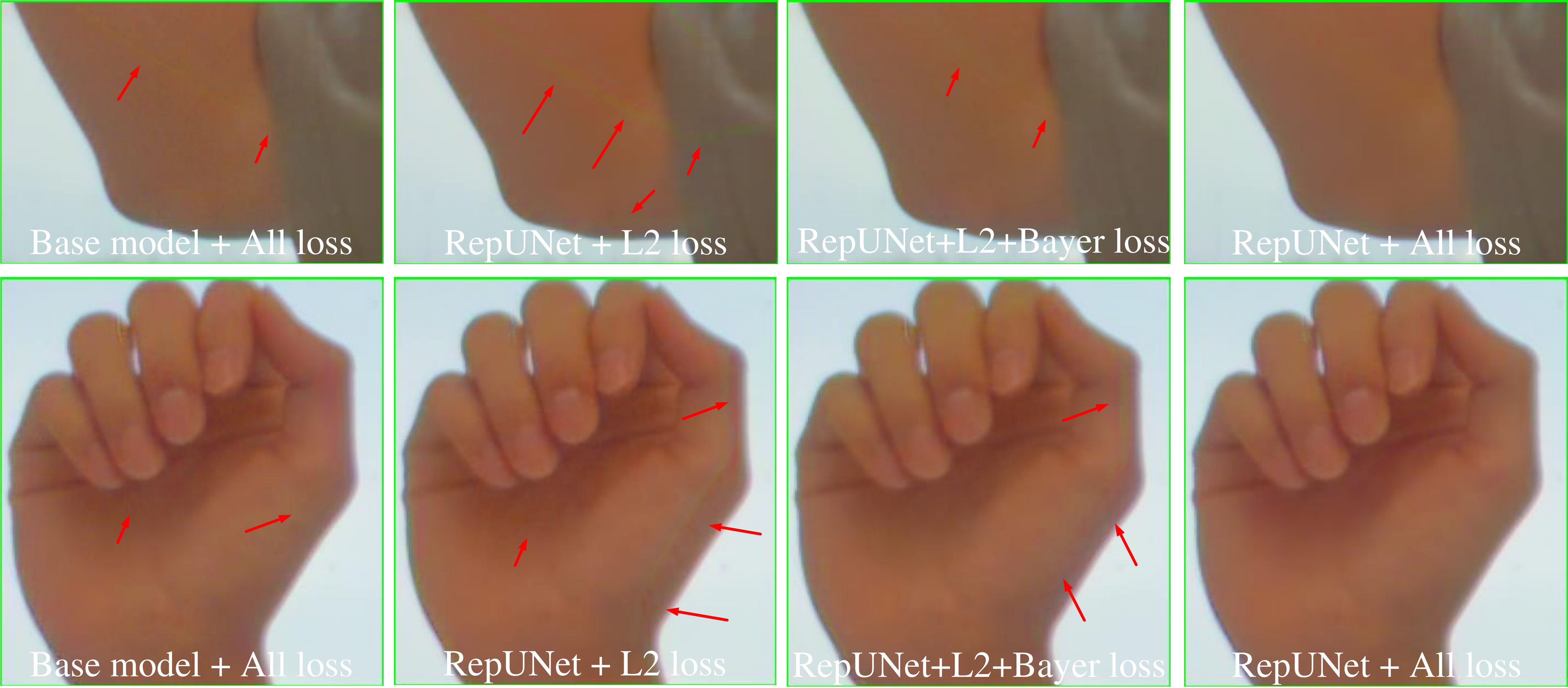}}  
    \caption {Visual results of RepUNet and its baseline variants. Combining these loss functions allows our model to produce top-notch results for motion and saturated areas effectively.}
    \vspace{-0.4cm}
    \label{loss_base}
\end{figure}

\begin{table*}[htb]
\centering
\caption{Reparameterization ablation results. The FLOPs and run times are measured on the raw image with a 4K resolution.}
\scalebox{1}{
\begin{tabular}{c|c|c|c|c|c|c|c|c|c|c|c}
\toprule
\multirow{2}{*}{{Method}} & \multirow{2}{*}{FLOPs} & \multirow{2}{*}{Params} & \multirow{2}{*}{Run Times} & \multicolumn{2}{c|}{All-Exposure} & \multicolumn{2}{c|}{Ratio=4} & \multicolumn{2}{c|}{Ratio=8} & \multicolumn{2}{c}{Ratio=16}\\
\cline{5-12}
 & & & & PSNR$\uparrow$ & $\Delta E$ $\downarrow$ & PSNR$\uparrow$	& $\Delta E$ $\downarrow$ & PSNR$\uparrow$	& $\Delta E$ $\downarrow$ & PSNR$\uparrow$	& $\Delta E$ $\downarrow$ \\
\midrule
{Base Model}        & 93.26G   & 0.82M  & 3.0 ms  & 39.7941	& 2.1202	& 39.8026	& 2.1036	& 40.5092	& 1.9790	& 39.0705	& 2.2780\\ 
% {DualUNet$_\textrm{tcb}$}     & 250.79G  & 2.16M  & 70.1 ms  & 40.5220	& 1.9573	& 40.4051	& 1.9746	& 41.3990	& 1.7978	& 39.7620	& 2.0995\\ 
{RepUNet} & 93.26G   & 0.82M  & 2.9 ms  & \textbf{40.5238}	& \textbf{1.9568}	& \textbf{40.4061}	& \textbf{1.9743}	& \textbf{41.4010}	& \textbf{1.7974}	& \textbf{39.7642}	& \textbf{2.0988}\\ 
\bottomrule
\end{tabular}}
\vspace{-0.3cm}
\label{re}
\end{table*}

\textbf{Loss functions.}
To test the effects of alignment-free and motion-aware short-exposure-first selection loss and Bayer loss, we set the $L2$ joint $L_{\textrm{ssim}}$ loss as the baseline and step-by-step modify the loss function combination. Tab. \ref{loss} and Fig. \ref{loss_base} show that adding the AMSS and Bayer loss steadily improves visual quality and quantitative results. RepUNet with joint loss achieves the best results, outperforming the baseline by 0.5 dB in PSNR and by 0.16 in $\Delta E$ on average. As Fig. \ref{loss_base} shows, alignment-free and motion-aware short-exposure-first selection loss (AMSS-Loss) effectively suppresses the ghosting artifacts (see columns 3 and 4). Meanwhile, our proposed Bayer loss can alleviate the color cast (see columns 2 and 3).

\textbf{Model reparameterization.}
Tab. \ref{re} presents the results for the base model and RepUNet. The RepUNet enjoys the same low complexity as the base model and shares even slightly higher reconstruction performance than RepUNet$_\textrm{tcb}$, which validates the effectiveness of our proposed TCB module. As can be seen, the enhanced models again obtain $0.7$dB consistent improvement on the PSNR index. This indicates that our TCB is a general drop-in replacement module for improving HDR performance without introducing additional inference costs.

\section{Conclusion}
In the paper, we proposed a Topological Convolution Block (TCB) for an efficient and lightweight HDR design that may be suitable for mobile devices. Based on the proposed TCB, we further designed RepUNet, aiming at balancing hardware efficiency and PSNR/SSIM indexes. Furthermore, we propose a novel computational photography-based pipeline for raw HDR image formation and construct a real-world raw HDR dataset -- RealRaw-HDR. Meanwhile, we designed a plug-and-play alignment-free and motion-aware short-exposure-first selection loss to mitigate ghost artifacts. Our empirical evaluation validates the effectiveness of the proposed SDR-HDR formation pipeline, as well as experiments show that our method achieves comparable performance to the state-of-the-art methods with less computational cost.

\section{Limitations and Discussion}
In this work, we propose an efficient and lightweight HDR network for dual-exposure HDR sensors. We have designed the TCB to prioritize computational efficiency, but real-world mobile deployments require additional hardware-specific adaptations. We will address this issue in future work through real-device testing. In addition, the proposed data synthesis pipeline only supports dual-exposure HDR sensor settings, and we will explore data synthesis pipelines for a wider range of triple-exposure HDR sensors in subsequent research. Additionally, we will focus on exploring more complex application scenarios such as flickering lights and large motion scenes.

\bibliographystyle{IEEEtran}
\bibliography{egbib}

\vspace{-2ex}
\begin{IEEEbiography}[{\includegraphics[width=1in,height=1.25in, clip,keepaspectratio]{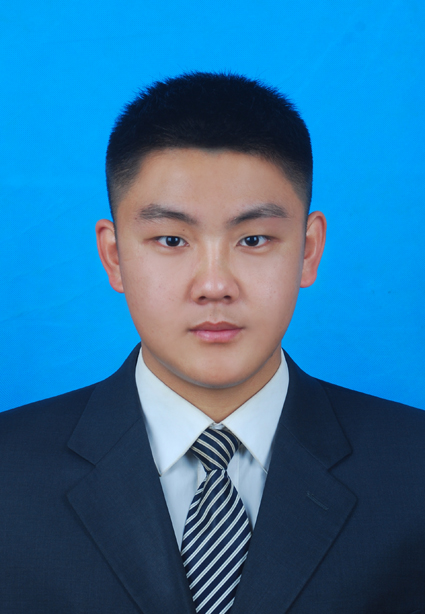}}]{Qirui Yang} received the M.E. degree in Control Engineering from University of Chinese Academy of Sciences, China, in 2021, and is currently working toward the PhD degree in the School of Electrical and Information Engineering, Tianjin University, Tianjin, China. His current research interests include computational photography, AI ISP, image enhancement, and restoration.
\end{IEEEbiography}

\vspace{-2ex}
\begin{IEEEbiography}[{\includegraphics[width=1in,height=1.25in, clip,keepaspectratio]{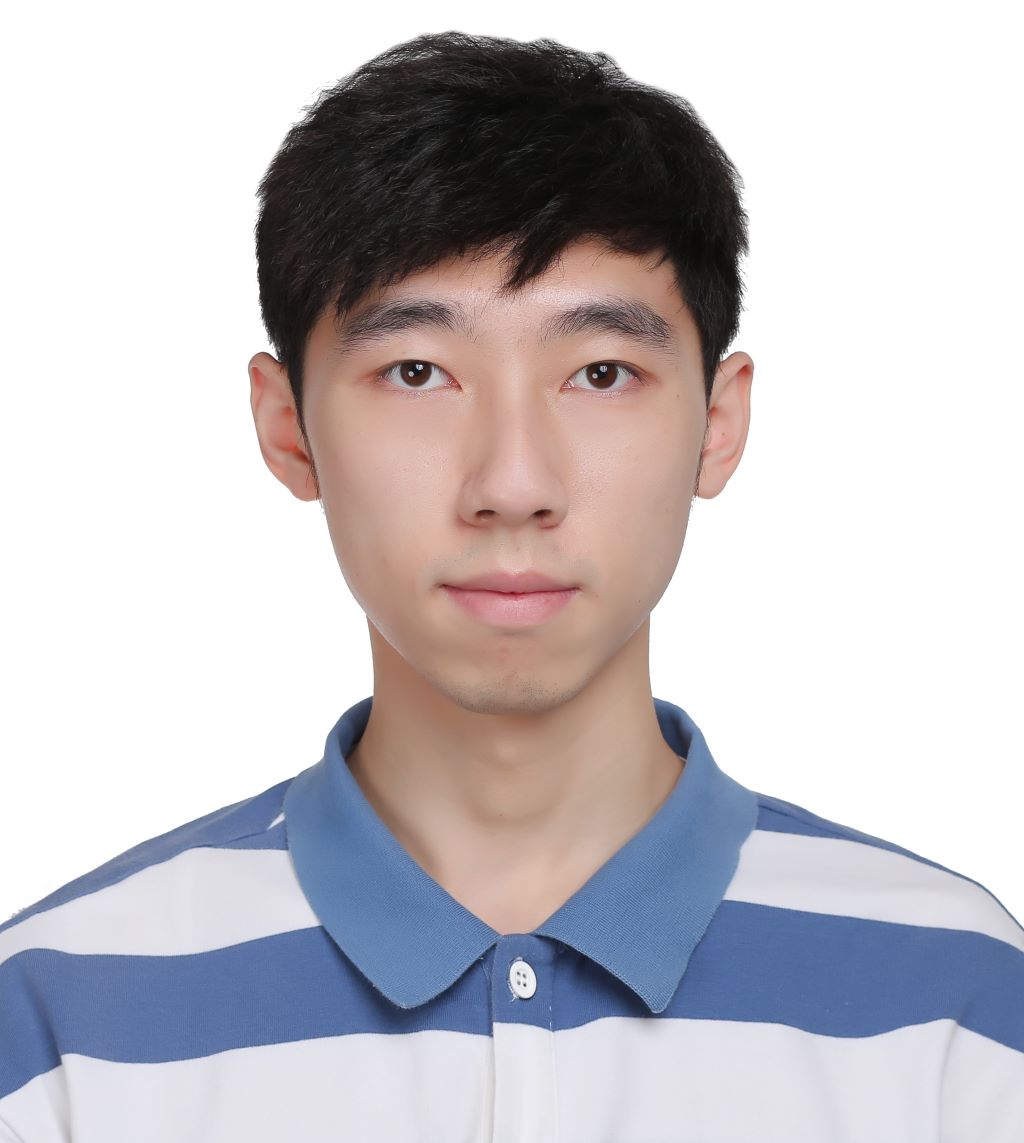}}]{Yihao Liu} is now a research scientist at Shanghai Artificial Intelligence Laboratory. He received the Bachelor’s degree in 2018 and the Ph.D. in 2023, both from the University of Chinese Academy of Sciences (UCAS). During his doctoral studies, he was affiliated with the Shenzhen Institutes of Advanced Technology (SIAT), Chinese Academy of Sciences. His research focuses on computer vision, particularly image/video restoration and enhancement techniques, and their real-world applications.
\end{IEEEbiography}

\vspace{-2ex}
\begin{IEEEbiography}[{\includegraphics[width=1in,height=1.25in, clip,keepaspectratio]{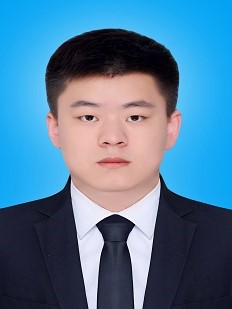}}]{Qihua Cheng} received the B.S. and M.E. degrees in Computer Engineering from Nanjing University of Science and Technology, China, in 2020, and is currently working in Shenzhen MicroBT Electronics Technology Co., Ltd, China. His current research interests include computational photography, AI ISP, image enhancement, and restoration.
\end{IEEEbiography}

\vspace{-2ex}
\begin{IEEEbiography}[{\includegraphics[width=1in,height=1.25in, clip,keepaspectratio]{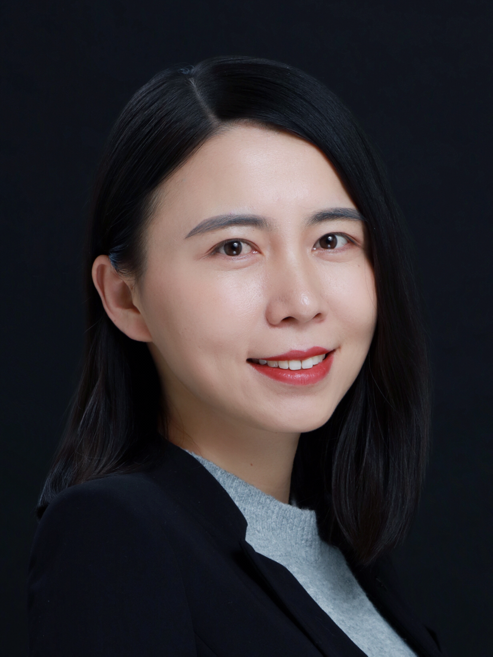}}]{Huanjing Yue} (Senior Member, IEEE) received the B.S. and Ph.D. degrees from Tianjin University, Tianjin, China, in 2010 and 2015, respectively. She was an intern with Microsoft Research Asia from 2011 to 2012 and from 2013 to 2015. She visited the Video Processing Laboratory, University of California at San Diego, from 2016 to 2017. She is currently an Associate Professor with the School of Electrical and Information Engineering, Tianjin University. Her current research interests include image processing and computer vision. She received the Microsoft Research Asia Fellowship Honor in 2013. She was selected into the Elite Scholar Program of Tianjin University in 2017.
\end{IEEEbiography}

\vspace{-2ex}
\begin{IEEEbiography}[{\includegraphics[width=1in,height=1.25in, clip,keepaspectratio]{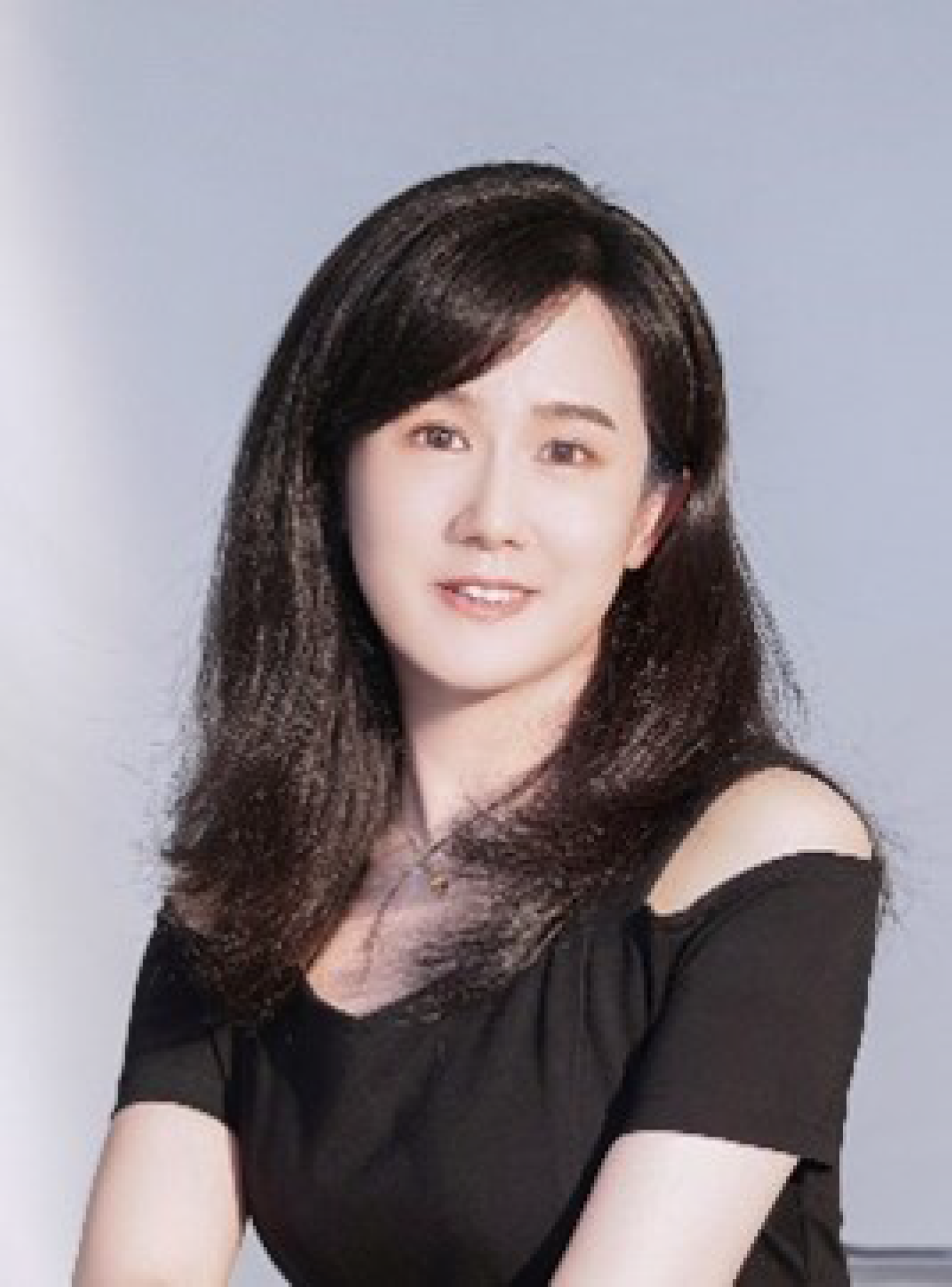}}]{Kun Li} (Senior Member, IEEE) received the B.E. degree from Beijing University of Posts and Telecommunications, Beijing, China, in 2006, and the master and Ph.D. degrees from Tsinghua University, Beijing, in 2011. She is currently a Professor with the College of Intelligence and Computing, Tianjin University, Tianjin, China. She was the recipient of the CSIG Shi Qingyun Award for Women Scientists. Her research interests include 3D reconstruction and AIGC.
\end{IEEEbiography}

\vspace{-2ex}
\begin{IEEEbiography}[{\includegraphics[width=1in,height=1.25in, clip,keepaspectratio]{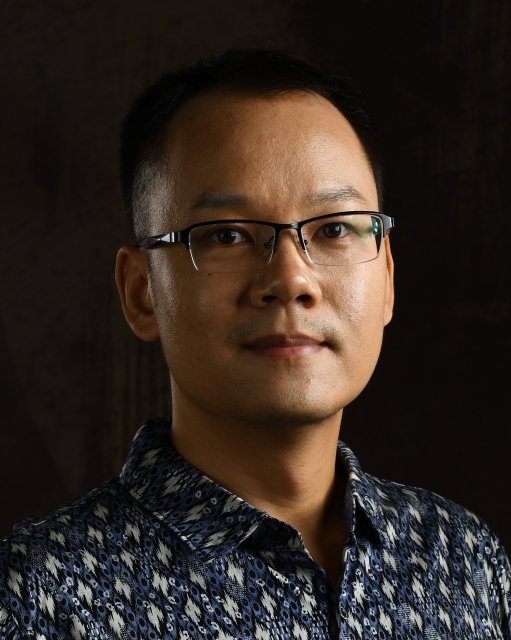}}]{Jingyu Yang} (Senior Member, IEEE) received the B.E. degree from the Beijing University of Posts and Telecommunications, Beijing, China, in 2003, and the Ph.D. degree (Hons.) from Tsinghua University, Beijing, in 2009. Since 2009, he has been a Faculty Member with Tianjin University, China, where he is currently a Professor with the School of Electrical and Information Engineering. He was with Microsoft Research Asia (MSRA), Beijing, in 2011, within the MSRAs Young Scholar Supporting Program, and with the Signal Processing Laboratory, École Poly technique Fédérale de Lausanne (EPFL), Lausanne, Switzerland, in 2012 and from 2014 to 2015. He has authored or coauthored over 180 high quality research papers (including dozens of IEEE TRANSACTIONS and top conference papers). His research interests include computer vision, computational photography, and 3D vision. As a coauthor, he got the Best 10\% Paper Award in IEEE VCIP 2016 and the Platinum Best Paper Award in IEEE ICME 2017. He was the Special Session Chair of the International Conference on Visual Communications and Image Processing in 2016 and the Area Chair of the International Conference on Image Processing in 2017. He was selected with the Program for New Century Excellent Talents in University (NCET) from the Ministry of Education, China, in 2011, the Tianjin Municipal Innovation Talent Promotion Program in 2015, and the National High-Level Youth Talent Program, China, in 2020. 
\end{IEEEbiography}

\vfill

\newpage

\newpage
\clearpage
\clearpage
% \setcounter{page}{1}

% \renewcommand\thesection{\Alph{section}}
% \appendix
\onecolumn

% \setcounter{section}{0}

% \maketitlesupplementary

%
\begin{center}
%	\vspace{-2mm}
	\Large\textbf{{Appendix}}\\
	% \vspace{-0.1cm}
\end{center}

In this Appendix, we provide additional results and analysis.

\section{Supplementary Materials}
\begin{figure*}[!ht]
    \centering
    \includegraphics[width=\textwidth]{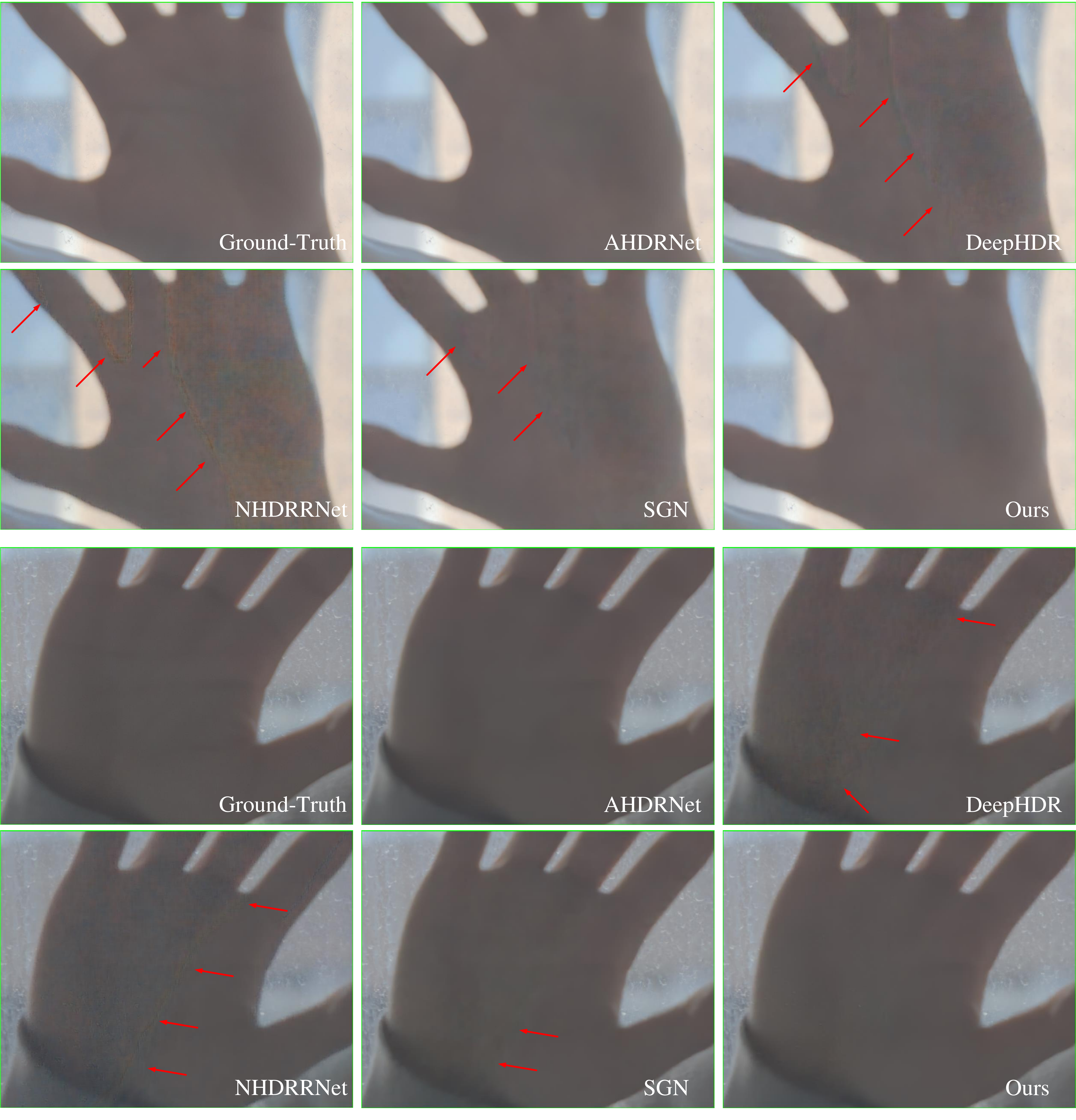}
    \caption{Visual comparison of state-of-the-art HDR reconstruction methods on the FUJI Raw dataset, captured with a Fujifilm GFX50S II camera containing real-world bracketed-exposure raw images.}
\label{com_4}
\end{figure*}

Fig. \ref{com_1} to Fig. \ref{com_4} show additional visual comparisons of our method with state-of-the-art HDR reconstruction methods. As shown in these figures, DeepHDR, NHDRRNet, and SGN all have obvious ghosting and color residuals, especially for moving color objects, and the reconstruction results will have severe ghosting. Our proposed RepUNet can reconstruct HDR images without ghosting and color bias, and we can effectively remove the noise in the reconstructed HDR. 

\begin{figure*}[htb]
    \centering
    \includegraphics[width=\textwidth]{Figure/sup_chen.pdf}
    \caption{Visual comparisons with the state-of-the-art methods on the actual HDR sensor raw dataset from Chen's dataset \cite{2021HDR}.}
\label{com_1}
\end{figure*}

\begin{figure*}[t]
    \centering
    \includegraphics[width=\textwidth]{Figure/more_fuji.pdf}
    \caption{Visual comparison of state-of-the-art HDR reconstruction methods on the FUJI Raw dataset, captured with a Fujifilm GFX50S II camera containing real-world bracketed-exposure raw images.}
\label{com_2}
\end{figure*}

\begin{figure*}[t]
    \centering
    \includegraphics[width=\textwidth]{Figure/sup_fujitest_2.pdf}
    \caption{Visual comparison of state-of-the-art HDR reconstruction methods on the FUJI Raw dataset, captured with a Fujifilm GFX50S II camera containing real-world bracketed-exposure raw images.}
\label{com_3}
\end{figure*}

\end{document}